\documentclass[twocolumn,epjc3]{svjour3}  

\RequirePackage{fix-cm}  
\journalname{Eur. Phys. J. A}

\usepackage{hyperref}
\usepackage[numbers,square,sort&compress]{natbib}

\usepackage{bebands}
\newcommand{\fnparityspace}{
  \label{fn:parity-space}
For odd-mass $p$-shell nuclei, as considered here, note
  that the ``natural'' parity, obtained with the lowest allowed filling of
  harmonic oscillator shells, as in a traditional ``$0\hw$'' shell model
  description, is negative parity.  An NCCI basis consisting of configurations with even
  numbers of oscillator excitations ($\Nex=0,2,\ldots$), and thus having an even
  $\Nmax$, yields the natural parity space, while an NCCI basis consisting of
  configurations with odd numbers of oscillator excitations ($\Nex=1,3,\ldots$), and
  thus odd $\Nmax$, yields the unnatural parity space.
}

\newcommand{\fnshellcoupling}{
  \label{fn:shell-coupling}
In the context of the shell model, where the strength of the spin-orbit
interaction controls the transition from $LS$ to $jj$ coupling, these schemes
are termed the \textit{weak coupling} and \textit{strong coupling} regimes,
respectively~\cite{inglis1953:p-shell}.
}

\newcommand{\fncollectivecoupling}{
  \label{fn:collective-coupling}
Here we encounter the terminology of weak \textit{vs.} strong coupling now in
the sense not of shell model angular momentum coupling schemes
(footnote~\ref{fn:shell-coupling}) but rather of collective motion, in which the
nucleus is presumed to factorize into a collective core and residual degrees of
freedom, typically the last odd ``uncoupled'' nucleon (see Sec.~1.8 of
Ref.~\cite{rowe2010:rowanwood} and Sec.~7.5 of
Ref.~\cite{rowe2010:collective-motion}).  In weak coupling, the residual degree
of freedom (in the present example, the last unpaired neutron spin), only
combines with the collective motion through angular momentum coupling to yield
the total angular momentum of the system, while, in strong coupling, the
nominally residual degree of freedom in fact fully participates in the intrinsic
wave function $\tket{\phi_K}$.  Thus the adiabatic rotational wave function
described in~(\ref{eqn:psi}) represents the strong coupling limit of collective
rotation.
}

\newcommand{\fnlanczostrick}{
  \label{fn:lanczos-trick}
The Lanczos trick, originally devised for evaluating strength
functions~\cite{whitehead1980:lanczos,mathews1985:lanczos-isnsm84,haxton2005:lanczos-em-response-28si},
may be used to obtain the decomposition of a wave function with respect to
eigenstates of any given Hermitian observable operator, such as the squared
angular momentum operators, to obtain angular momentum
decompositions~\cite{johnson2015:spin-orbit}, or a more general group's Casimir
operator, to obtain a decomposition into irreps of this
group~\cite{gueorguiev2000:fp-su3-breaking}.  It is necessary only to take the
calculated wave function and use it as the new pivot vector for a Lanczos
diagonalization of the $\Lvec^2$ operator, $\Svec^2$ operator, or other operator
of interest.
}
  
\newcommand{\fnexptassignment}{
  \label{fn:expt-assignment}
  The experimental level at $3.4\,\MeV$ excitation energy in
                  $\isotope[11]{Be}$, which is only identified as
                  $(3/2^-,3/2^+)$ in Ref.~\cite{npa2012:011}, is shown as
                  $3/2^-$ in Fig.~\ref{fig:levels-11be}(a), consistent with
                  the rotational analysis of
                  Refs.~\cite{vonoertzen1997:be-alpha-rotational,bohlen2008:be-band},
                  where it is taken as a $K^P=3/2^-$ band head.  See discussion
                  in footnote~6 of Ref.~\cite{caprio2019:bebands-ntse18}.
}

\newcommand{\fnsevenbehint}{
  \label{fn:7be-hint}
A hint of the excited band structure may already be found in Fig.~10 of
                  Ref.~\cite{caprio2015:berotor-ijmpe}, where enhanced
                  transitions may be seen from the yrast $9/2^-$ and $11/2^-$
                  states to high-lying $5/2^-$ and $7/2^-$ states.  However, the
                  calculations of 
                  Refs.~\cite{caprio2013:berotor,maris2015:berotor2,maris2019:berotor2-ERRATUM,caprio2015:berotor-ijmpe}
                  were based on the less rapidly convergent JISP16 and
                  \nnloopt{} interactions, and only carried to $\Nmax=10$, so
                  the excited band members lay at higher excitation energy, obscured in a
                  region of higher level density, in these earlier
                  calculations.
}

\newcommand{\fnpmam}{
  \label{fn:pmam}
The expectation value of $\tbracket{\Lvec^2}$, and thus the effective $\bar{L}$,
can always be recovered from the full angular momentum decomposition by $L$, and
thus provides no new information relative to this full decomposition (and
similarly for the effective spin angular momenta).  However,
$\tbracket{\Lvec^2}$ is the expectation value of a rotational scalar two-body
operator, a standard class of observables to extract from NCCI wave functions, and can
be computed much more efficiently than the full decomposition, which requires
further Lanczos diagonalizations.
}

\newcommand{\fnninebetermination}{
\label{fn:9be-termination}
The yrast $15/2^+$ and $19/2^+$ states were considered in
Refs.~\cite{maris2015:berotor2,caprio2015:berotor-ijmpe} as
possible extended members of the $K^P=1/2^+$ band, on the
basis of strong $E2$ transitions to the $11/2^+_1$ and
$13/2^+_1$ band members, but these states lie at energies
above what would be expected from the rotational energy
formula.  These states would now seem more likely to be
members of a higher-lying excited rotational band with strong
interband transitions, as discussed in detail for the
illustrative case of $\isotope[7]{Be}$ in Sec.~\ref{sec:7be}.
}

\newcommand{\fnamadmixture}{
\label{fn:am-admixture-EXTRACTED}
When interpreting small values of $\bar{L}$ or $\bar{S}$, we must keep in
                  mind the nonlinear relationship entering into the definition
                  of the effective angular momentum.  A small admixture of
                  higher spin into an $S=0$ state can have an outsized effect on
                  $\bar{S}$, \textit{e.g.}, an effective $\bar{S}\approx0.2$ is
                  obtained with only a $10\%$ admixture of $S=1$.
}

\newcommand{\ifproofpre}[2]{#1}

\begin{document}

\title{Probing \textit{ab initio} emergence of nuclear rotation}

\author{
  Mark A. Caprio\thanksref{addr:nd}
  \and
  Patrick J. Fasano\thanksref{addr:nd}
  \and
  Pieter Maris\thanksref{addr:isu}
  \and
  Anna E. McCoy\thanksref{addr:triumf}
  \and
  James P. Vary\thanksref{addr:isu}
}

\institute{
  Department of Physics, University of Notre Dame, Notre Dame, Indiana 46556-5670, USA \label{addr:nd}
  \and
  Department of Physics and Astronomy, Iowa State University, Ames, Iowa
  50011-3160, USA \label{addr:isu}
  \and
  TRIUMF, Vancouver, British Columbia V6T~2A3, Canada \label{addr:triumf}
}

\date{Received: date / Accepted: date}

\maketitle

\begin{abstract}
%
%
Structural phenomena in nuclei, from shell structure and clustering
to superfluidity and collective rotations and vibrations, reflect emergent
degrees of freedom.  \textit{Ab initio} theory describes
nuclei directly from a fully microscopic formulation.
%
%
We can therefore look to \textit{ab initio} theory as a means of exploring the
emergence of effective degrees of freedom in nuclei.
%
%
For the illustrative case of emergent rotational bands in the $\isotope{Be}$
isotopes, we establish an understanding of the underlying oscillator space and
angular momentum (orbital and spin) structure.
%
%
We consider no-core configuration interaction (NCCI) calculations for
$\isotope[7,9,11]{Be}$ with the Daejeon16 internucleon interaction.
%
%
Although shell model or rotational degrees of freedom are not assumed in the
\textit{ab initio} theory, the NCCI results are suggestive of the emergence of
effective shell model degrees of freedom ($0\hw$ and $2\hw$ excitations) and
$LS$-scheme rotational degrees of freedom, consistent with an Elliott-Wilsdon
$\grpsu{3}$ description.
%
%
%
These results provide some basic insight into the connection between emergent
effective collective rotational and shell model degrees of freedom in these
light nuclei and the underlying \textit{ab initio} microscopic description.
\end{abstract}



\section{Introduction}
\label{sec:intro}

Nuclei are quantum many body systems where the structural phenomena and
spectroscopic features characteristically reflect emergent degrees of freedom,
from shell structure and clustering to superfluidity and collective rotations
and vibrations.  These degrees of freedom are traditionally the domain of
phenomenological
models~\cite{eisenberg1987:v1,iachello1987:ibm,talmi1993:shell-ibm,casten2000:ns,rowe2010:rowanwood},
yet their description may be placed on a more fundamental footing if viewed in
terms of effective theories built on a microscopic description.  Indeed, the
emergent phenomena of nuclear structure may be viewed as simply the topmost tier
in a tower of effective theories of nuclear physics beginning at the
subnucleonic level~\cite{papenbrock2011:eft-rotation,furnstahlXXXX:edf-effective-epja}.

\textit{Ab initio} nuclear theory attempts a direct description explicitly from
the fully microscopic formulation of the many-body system in terms of nucleons
and their free-space interactions.  An accurate treatment of correlations can be
computationally challenging, but \textit{ab initio} theory now reproduces
signatures of emergent phenomena, including
clustering~\cite{pieper2004:gfmc-a6-8,neff2004:cluster-fmd,maris2012:mfdn-ccp11,romeroredondo2016:6he-correlations,navratil2016:ncsmc}
and
rotation~\cite{caprio2013:berotor,maris2015:berotor2,maris2019:berotor2-ERRATUM,caprio2015:berotor-ijmpe},
primarily in light nuclei. We can look therefore to \textit{ab initio} theory as
a means of exploring the emergence of effective degrees of freedom from a
microscopic foundation, and understanding their place within the full
description of nuclear properties and spectroscopy.

To explore the physical structure of emergent rotation in some of the lightest
nuclei, and to gain some insight into the nature of the relevant effective
degrees of freedom, we consider here a few illustrative ``case studies'' of
rotational bands in \textit{ab initio} no-core configuration interation
(NCCI)~\cite{barrett2013:ncsm}, or no-core shell model (NCSM), calculations of
the odd-mass $\isotope{Be}$ isotopes, specificially, $\isotope[7,9,11]{Be}$.

Some of these rotational bands (or portions thereof) have been studied before in
NCCI
calculations~\cite{caprio2013:berotor,maris2015:berotor2,maris2019:berotor2-ERRATUM,johnson2015:spin-orbit,caprio2015:berotor-ijmpe}.
However, these earlier calculations, which were based on internucleon
interactions such as JISP16~\cite{shirokov2007:nn-jisp16} and
\nnloopt{}~\cite{degregorio2017:hf-tda-nnloopt}, suffered from a significant
limitation: excited rotational bands were relatively poorly converged in the
many-body calculations, lying at much higher excitation energy than they would
either in a more completely converged calculation for those interactions or,
indeed, in experiment.  In practice, this meant that portions of the bands,
especially at lower angular momentum, lay in regions of the spectrum where the
calculated level density was high, and mixing or fragmentation consequently
tended to obscure these bands.

Here we make use of the much softer Daejeon16 internucleon
interaction~\cite{shirokov2016:nn-daejeon16}, which permits more rapidly
convergent NCCI calculations.  This interaction is obtained starting from the
classic Entem-Machleidt \nthreelo{} chiral perturbation theory
interaction~\cite{entem2003:chiral-nn-potl}, which is then softened via a
similarity renormalization group transformation and subsequently adjusted via a
phase-shift equivalent transformation to yield an accurate description of light
nuclei with $A\leq16$.

In cases where only fragmented precursors of excited rotational bands were
obtained in previous calculations, these bands are now found low in the
excitation spectrum, and approaching their converged energies.  Improved
convergence means not only that the rotational energies themselves are more
accurately described, but that the band structure itself becomes clearer.  The
cleaner rotational spectrum, comparatively free from fragmentation, permits
easier interpretation and understanding of the electric quadrupole ($E2$)
transition spectroscopy.

Improved convergence also, in principle, allows more meaningful identification
of experimentally observed counterparts to the rotational states.  However,
doing so requires detailed understanding of the energy
convergence and, for the less completely
converged energies, likely still entails some form of basis
extrapolation~\cite{bogner2008:ncsm-converg-2N,forssen2008:ncsm-sequences,maris2009:ncfc,coon2012:nscm-ho-regulator,furnstahl2014:ir-expansion,furnstahl2015:ir-extrapolation-cc-16o}.
Our focus here will thus primarily be restricted to the nature of the emergent
structure arising in solving the many-body problem, rather than detailed
comparison with experiment (\textit{e.g.}, Refs.~\cite{caprio2019:bebands-ntse18,chen20xx:11be-xfer}).

Beyond simply analyzing the spectroscopic signatures of emergent phenomena
appearing in \textit{ab initio} calculations, we can make use of microscopic
wave functions obtained in these calculations to directly probe for structural
insight.  In the following discussions, we examine the decompositions of the
wave functions in terms of oscillator excitations (\textit{i.e.}, ``$0\hw$'' and
``$2\hw$ or higher contributions) and spin and orbital angular momentum
contributions.  We build here on Johnson's
analysis~\cite{johnson2015:spin-orbit} of the angular momentum structure of
rotational states.

In characterizing effective theories of emergent nuclear phenomena, we consider
not only emergent degrees of freedom, but also emergent symmetries.
Dynamical
symmetries~\cite{gellmann1962:su3-flavor,wybourne1974:groups,barut1977:group-repn,iachello1979:dynsymm}
in general can be responsible for the emergence of simple patterns in the
behavior of complex systems.  In the traditional shell model, Elliott's
$\grpsu{3}$ dynamical
symmetry~\cite{elliott1958:su3-part1,elliott1958:su3-part2,elliott1963:su3-part3,elliott1968:su3-part4}
provides a mechanism for the emergence of rotation, as arising naturally within
the $\grpsu{3}$ irreducible representations (irreps) preferred by a
quadrupole-quadrupole interaction.  The symplectic group $\grpsptr$, which
contains Elliot's $\grpsu{3}$ as a subgroup, and the multishell symplectic shell
model associated with this group, has then been proposed as providing a microscopic
formulation of the unified collective
model~\cite{rosensteel1977:sp6r-shell,rowe1985:micro-collective-sp6r,rowe2016:micsmacs}.
Wave functions obtained in \textit{ab initio} calculations have indeed been
found to receive strong contributions from specific dominant $\grpu{3}$ or
$\grpsptr$ symmetry
components~\cite{dytrych2007:sp-ncsm-evidence,dytrych2007:sp-ncsm-dominance,draayer2012:sa-ncsm-qghn11,dytrych2013:su3ncsm,launey2016:sa-nscm}.
We shall therefore comment, at least briefly, on how the rotational
structures considered here can relate to such symmetries.

In the following explorations, we begin with $\isotope[9]{Be}$
(Sec.~\ref{sec:9be}).  The low-lying rotational bands in both the natural
(negative) parity and unnatural (positive) parity spaces provide particularly
clean illustrations of the angular momentum structure of the rotational bands
and the implications of Elliott $\grpsu{3}$ dynamical symmetry.

Then $\isotope[11]{Be}$ (Sec.~\ref{sec:11be}) provides an example of the
coexistence of $0hw$ and $2\hw$ rotational bands within the same spectrum, and
thus of rotation outside the effective space of the $0\hw$ shell model.

Finally, $\isotope[7]{Be}$ (Sec.~\ref{sec:7be}) introduces the qualitatively
distinct situation in which the $0\hw$ ground state band is connected by strong
$E2$ transitions to a $2\hw$ excited band.  This excitation may be understood in
terms of $\grpsptr\supset\grpu{3}$ dynamical
symmetry~\cite{mccoy2018:diss,mccoyXXXX:spfamilies}.  In the macroscopic
limit, it would be construed as representing the excitation of an effective giant
quadrupole degree of freedom.

However, we first (Sec.~\ref{sec:overview}) lay out the excitation spectra
obtained in \textit{ab initio} NCCI calculations for all three nuclei.  In
preparation for the following discussions, we review the basic spectroscopic
properties of nuclear rotations, as well as stress the importance of
considering convergence (with respect to the basis for the NCCI calculation) in
interpreting any such results.

\section{Rotational signatures and overview of calculations}
\label{sec:overview}

\subsection{Rotational spectra}
\label{sec:overview:rotation}

Rotational bands are commonly recognized (whether in experiment or in
calculations) by energies following the rotational formula, relative $E2$
strengths following the rotational formulas (Alaga rules)~\cite{alaga1955:branching}, and enhanced $E2$
strengths overall.
In terms of the physical structure of the rotational band members, these
features arise from a shared intrinsic structure, combined with a different
overall rotational motion, for the different members of the same rotational band.  That is, in ideal
rotation, there is an adiabatic separation of
degrees of freedom, between a rotational intrinsic state and the collective
rotational coordinates (Euler angles).  In the case of an axially symmetric
intrinsic state $\tket{\phi_K}$, the angular momentum is characterized by a definite projection
$K$ onto the intrinsic symmetry axis, and the rotational band members have
angular momenta $J\geq K$, with wave functions~\cite{rowe2010:collective-motion}
\begin{multline}
\label{eqn:psi}
\tket{\psi_{JKM}}\propto \int d\vartheta\,\bigl[ ~
  \scrD^J_{MK}(\vartheta)
  \,
 \tket{\phi_K;\vartheta}
~ \\ + ~
 (-)^{J+K}
 \scrD^J_{M,-K}(\vartheta)
 \,
 \tket{\phi_{\bar{K}};\vartheta}
\bigr],
\end{multline}
where the Wigner $\scrD$ function constitutes the wave function in the
collective rotational Euler angle coordinate $\vartheta$, while
$\tket{\phi_K;\vartheta}$ represents the intrinsic state after rotation by Euler
angles $\vartheta$.  The second term, involving the conjugated intrinsic state
$\tket{\phi_{\bar{K}}}$ with angular momentum projection $-K$ along the
intrinsic symmetry axis, arises to preserve symmetry under rotation by an angle
$\pi$ about an axis perpendicular to the symmetry axis.

Band members are then expected to have energies following the rotational formula
\begin{equation}
\label{eqn:E-rotational-no-coriolis}
  E(J)=E_0+AJ(J+1),
\end{equation}
where the rotational energy constant
$A\equiv\hbar^2/(2\cal{J})$ is inversely related to the moment of inertia
$\cal{J}$ of the rotational intrinsic state, and the intercept parameter
$E_0=E_K-AK^2$ is related to the energy $E_K$ of the rotational intrinsic
state.  However, for $K=1/2$ bands, the Coriolis contribution to the kinetic
energy significantly modifies the energies, leading to an energy staggering
which is given, in first-order perturbation theory, by
\begin{equation}
\label{eqn:E-rotational-coriolis}
E(J)=E_0+A\bigl[J(J+1)+
a(-)^{J+1/2}(J+\tfrac12)
\bigr],
\end{equation}
where the Coriolis decoupling parameter $a$ depends upon the structure of the
rotational intrinsic state.

\begin{figure*}
\begin{center}
\includegraphics[width=\ifproofpre{0.90}{1}\hsize]{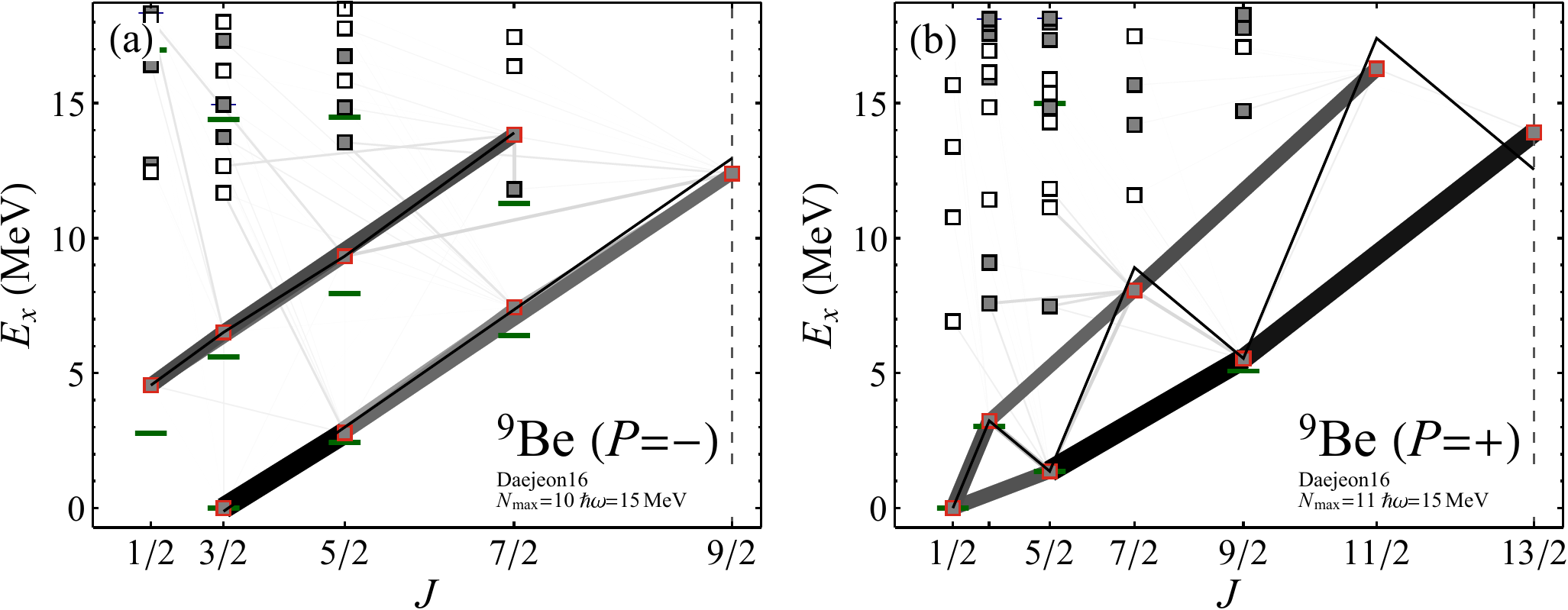}
\end{center}
\caption{\textit{Ab initio} calculated energy spectra for $\isotope[9]{Be}$
  (a)~negative and (b)~positive parity, obtained for the Daejeon16 interaction.
  Calculations are with basis truncations $\Nmax=10$ and $11$, respectively, and
  oscillator basis parameter $\hw=15\,\MeV$.  Experimental energies (green
  horizontal lines) are shown for comparison (see text).  Rotational band
  members are highlighted (red squares), and rotational energy fits are
  indicated by lines.  The $J$-decreasing $E2$ transitions originating from
  these rotational band members are shown (specifically, transitions with
  $J_f<J_i$ or with $J_f=J_i$ and $E_f<E_i$), where the line thickness (and
  shading) is directly proportional to the $B(E2)$ strength.  States are
  approximately classified as $0\hw$ (filled symbols) or $2\hw$ (open symbols)
  for natural parity, or similarly $1\hw$ and $3\hw$ for unnatural parity,
  classified by the dominant oscillator configuration (see
  Sec.~\ref{sec:9be:oscillator}), and states with isospin $T>\abs{T_z}$ are
  indicated by tick marks.  The maximal valence angular momentum, \textit{i.e.}, the largest which can be
  constructed in the $0\hw$ or $1\hw$ space, respectively, is indicated by the
  vertical dashed line.  Excitation energies are taken separately within each
  parity.
\label{fig:levels-9be}
}
\end{figure*}
\begin{figure*}
\begin{center}
\includegraphics[width=\ifproofpre{0.90}{1}\hsize]{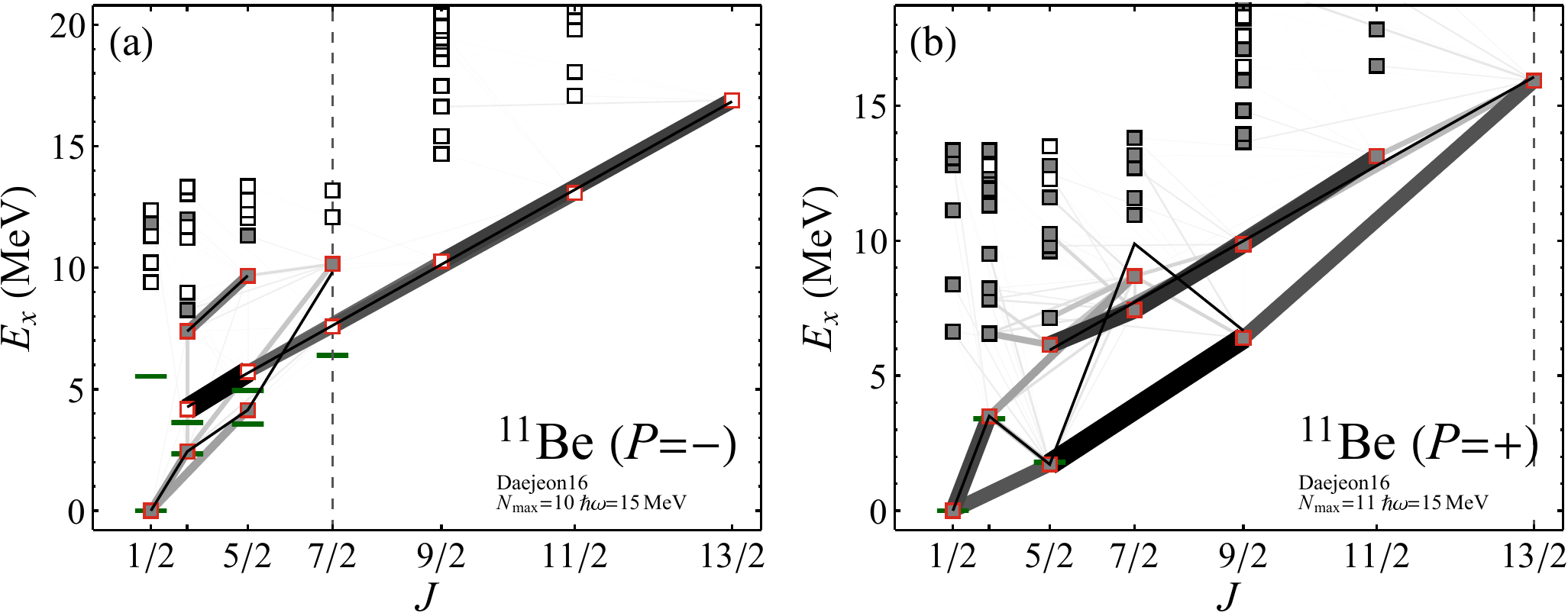}
\end{center}
\caption{\textit{Ab initio} calculated energy spectra for $\isotope[11]{Be}$
  (a)~negative and (b)~positive parity, obtained for the Daejeon16 interaction,
  with basis truncations $\Nmax=10$ and $11$, respectively.  See
  Fig.~\ref{fig:levels-9be} caption for further description of figure contents
  and labeling.
\label{fig:levels-11be}
}
\end{figure*}
\begin{figure}
\begin{center}
\includegraphics[width=\ifproofpre{0.90}{0.5}\hsize]{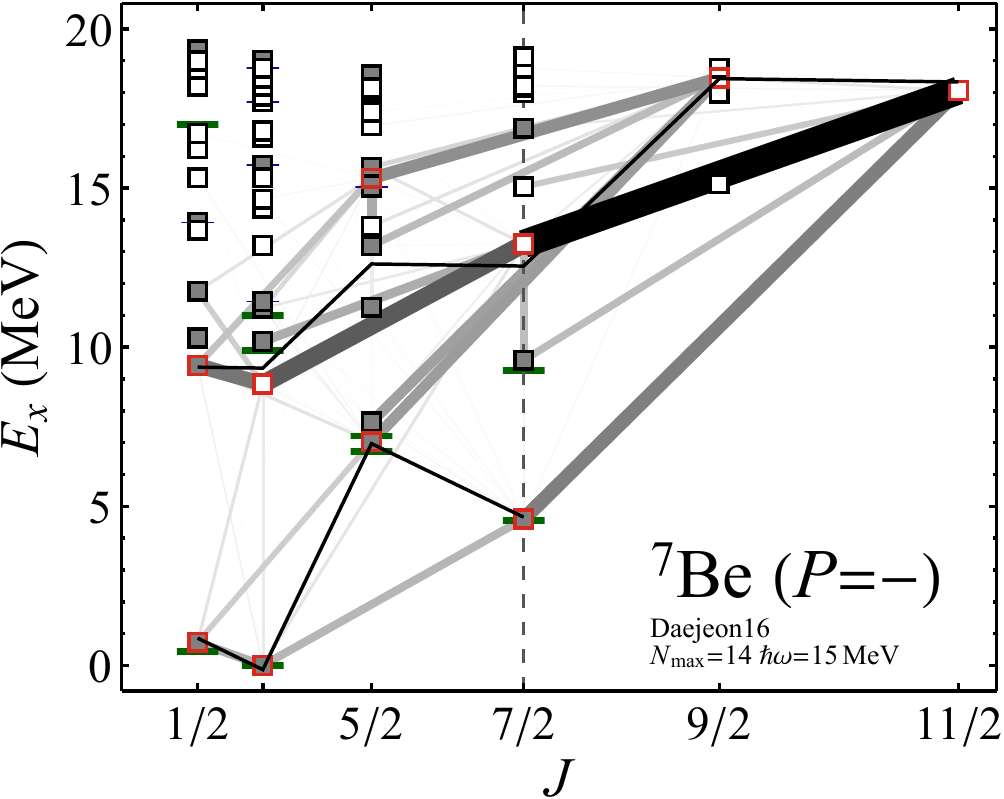}
\end{center}
\caption{\textit{Ab initio} calculated energy spectra for $\isotope[7]{Be}$
  negative parity, obtained for the Daejeon16 interaction, with basis truncation
  $\Nmax=14$.  See Fig.~\ref{fig:levels-9be} caption for further description of
  figure contents and labeling.
\label{fig:levels-7be}
}
\end{figure}

The excitation spectra obtained in NCCI calculations for the odd-mass isotopes
$\isotope[7,9,11]{Be}$ are shown in
Figs.~\ref{fig:levels-9be}--\ref{fig:levels-7be}, with energies against an
angular momentum axis scaled as $J(J+1)$, as appropriate for rotational energy
analysis.  The states identified with the rotational bands discussed in the
following are highlighted (red outlines).  The $E2$ transitions from these
levels are shown, specifically, for angular-momentum decreasing transitions
originating from band members.  The line thicknesses (and
shadings) indicate transition strengths.  Further details of the calculations
are defined below.

For the bands in Figs.~\ref{fig:levels-9be}--\ref{fig:levels-7be}, the energies
expected from the rotational energy
relations~(\ref{eqn:E-rotational-no-coriolis})
or~(\ref{eqn:E-rotational-coriolis}) are shown as best fit lines.  [For the
  $K=1/2$ bands, the three parameters in~(\ref{eqn:E-rotational-coriolis}) are
  simply determined to match the calculated energies of the three lowest-energy
  band members.]
\begin{figure}
\begin{center}
  \includegraphics[width=\ifproofpre{0.90}{0.5}\hsize]{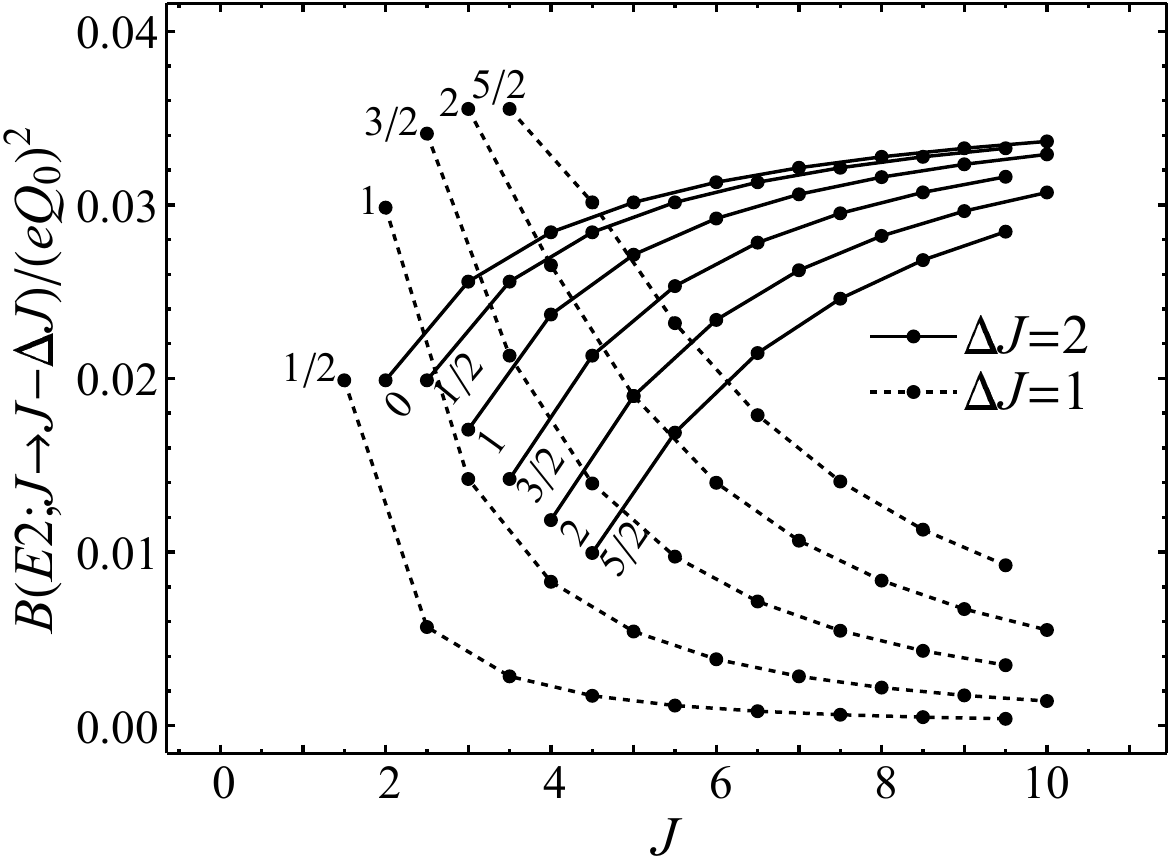}
\end{center}
\caption{Rotational Alaga rule predictions for $B(E2)$ strengths within a
  rotational band, normalized to the square of the intrinsic quadrupole moment
  $Q_0$, shown separately for $\Delta J=2$ (solid curves) and $\Delta J=1$
  (dotted curves) transitions.  Curves are shown for bands with $0\leq K \leq
  5/2$, as indicated.
  \label{fig:rotor-e2}
}
\end{figure}
\begin{figure}
\begin{center}
  \includegraphics[width=\ifproofpre{0.90}{0.5}\hsize]{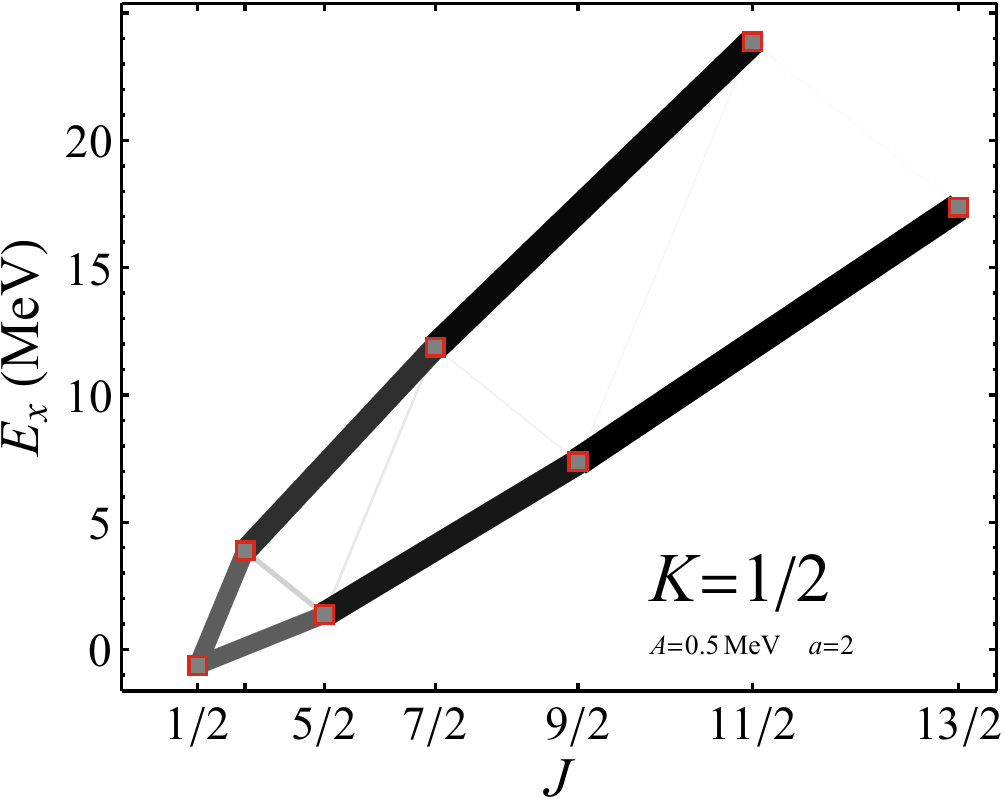}
\end{center}
\caption{Rotational predictions for $B(E2)$ strengths within an ideal $K=1/2$
  band.  The line thickness (and shading) is directly proportional to the
  $B(E2)$ strength.
  \label{fig:rotor-e2-network-k12}
  }
\end{figure}

A common criterion for identifying rotational band members is that, loosely
speaking, $E2$ strengths between rotational band members are expected to be
``enhanced''.  The $E2$ matrix elements within a rotational band follow the
Alaga relations, and thus all $E2$ matrix elements
within a band should follow a pattern of \textit{relative} intensities given by
Clebsch-Gordan coefficients.  Then, the overall \textit{scale} of the
intensities is determined by the $E2$ matrix element within the intrinsic state
or, equivalently, by the intrinsic quadrupole moment $Q_0$.  (Similar relations
apply to interband transitions between the same two rotational bands, with the
overall strength given by an interband intrinsic matrix element.)
For $B(E2)$ strengths (\textit{i.e.}, reduced transition probabilities) within a
band, the rotational relation becomes
\begin{equation}
\label{eqn:BE2-rotational}
B(E2;J_i\rightarrow J_f)=\frac{5}{16\pi} \tcg{J_i}{K}{2}{0}{J_f}{K}^2 (eQ_0)^2.
\end{equation}
This relation yields the strengths in Fig.~\ref{fig:rotor-e2}, where
different curves apply depending upon $K$ for the band, which, for an ideal
$K=1/2$ band, gives the transition pattern illustrated in
Fig.~\ref{fig:rotor-e2-network-k12}.

Consequently, $E2$ transitions within a band are enhanced to the extent that the
intrinsic matrix element is larger than the typical scale of $E2$ matrix
elements between arbitrary states not within a band.  This is commonly the case,
as rotation is associated with quadrupole deformation.  Even so, it is worth
keeping in mind that not all transitions within a rotational band are expected
to be ``strong'', if they are suppressed by the Clebsch-Gordan coefficient,
\textit{e.g.}, for the $K=1/2$ band (Fig.~\ref{fig:rotor-e2-network-k12}), the
$\Delta J=1$ transitions (with the exception of the $3/2\rightarrow1/2$ band
head transition) are highly suppressed relative to the $\Delta J=2$ transitions.

\subsection{Convergence of rotational observables}
\label{sec:overview:convergence}

The NCCI approach is based upon expressing the nuclear many-body system in terms
of a basis of antisymmetrized products (Slater determinants) of harmonic
oscillator single-particle states.  The many-body Hamiltonian is represented as
a matrix in terms of this basis, and the energy eigenvalues and wave functions
are obtained by solving the (large) matrix eigenproblem which ensues.

Calculations must, of course, be carried out in a finite, truncated basis, and
the results depend upon this truncation.  Each basis state represents a
configuration of nucleons distributed over oscillator shells, and is thus
characterized by the number $\Nex$ of oscillator excitations above the lowest
Pauli-allowed filling of oscillator shells.  The basis is commonly constrained
by limiting the number of excitations to $\Nex\leq\Nmax$.  (Results obtained
with a truncated basis also depend upon the underlying oscillator length
scale~\cite{suhonen2007:nucleons-nucleus} defining the basis, given by the basis
parameter $\hw$.)  However, as $\Nmax$ increases, results converge towards those
which would be obtained in the full, untruncated many-body space, and thus also
must become independent of the basis parameter $\hw$.

The results shown in Figs.~\ref{fig:levels-9be}--\ref{fig:levels-7be} are
obtained in spaces truncated to $\Nmax=10$ or $11$ for $\isotope[9,11]{Be}$
(Figs.~\ref{fig:levels-9be}--\ref{fig:levels-11be}) and $\Nmax=14$ for
$\isotope[7]{Be}$.\footnote{\fnparityspace}  These calculations were
obtained based on the Daejeon16 interaction plus Coulomb interaction, with
oscillator basis parameter $\hw=15\,\MeV$ (roughly corresponding to the
variational minimum energy), using the $M$-scheme NCCI code
MFDn~\cite{maris2010:ncsm-mfdn-iccs10,aktulga2013:mfdn-scalability,shao2018:ncci-preconditioned}.
Initial results from the present calculations were included in
Refs.~\cite{caprio2019:bebands-ntse18,chen20xx:11be-xfer}.
\begin{figure*}
\begin{center}
  \includegraphics[width=\ifproofpre{0.90}{0.95}\hsize]{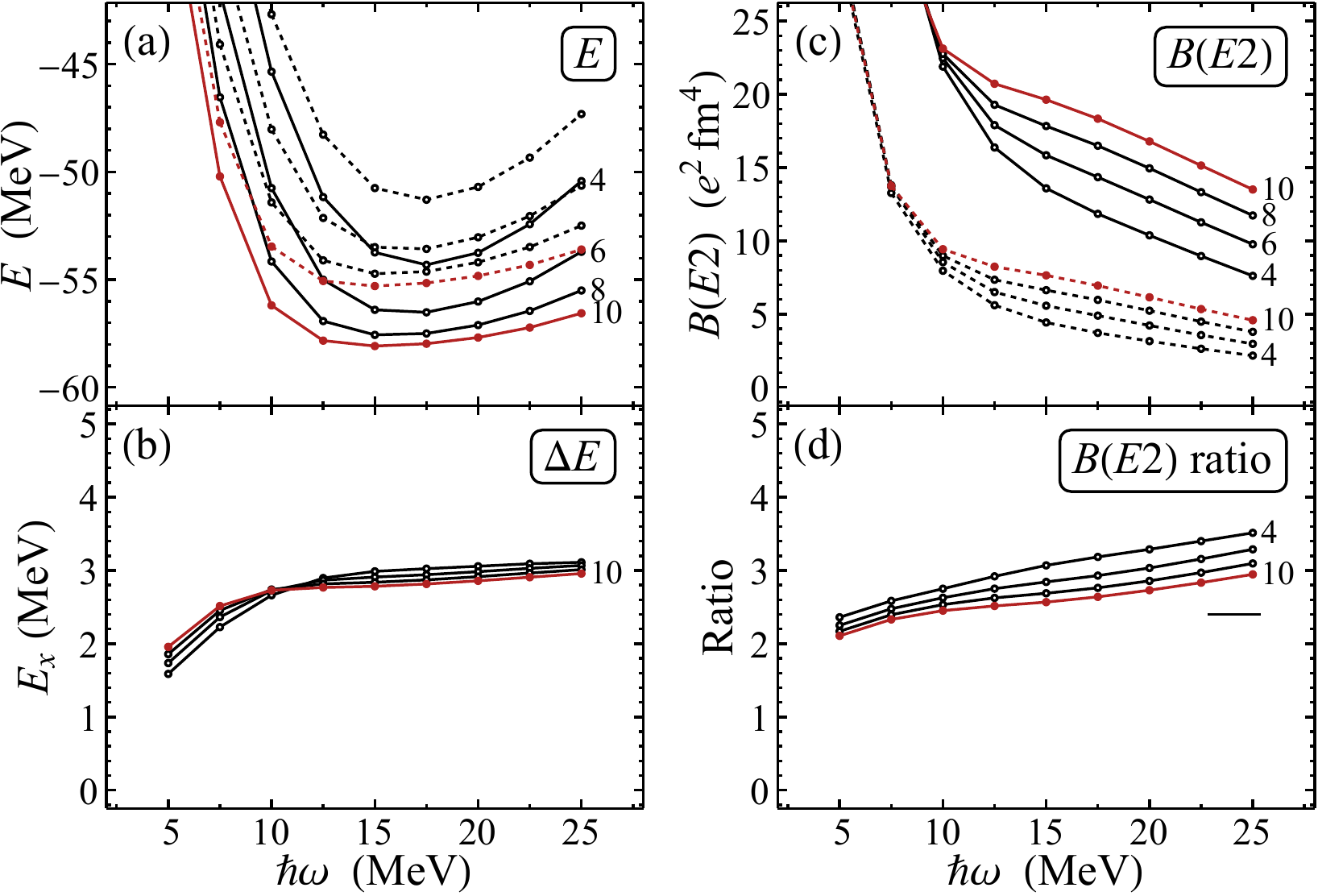}
\end{center}
\caption{Convergence of calculated energy and transition observables for
  $\isotope[9]{Be}$~(top) and corresponding relative observables~(bottom):
  (a)~Energies of the $3/2^-_1$ ground state (solid curves) and $5/2^-_1$
  rotational band member (dashed curves).  (b)~The energy
  difference $E(5/2^-)-E(3/2^-)$. (c)~Transition strengths
  $B(E2;5/2^-\rightarrow3/2^-)$ (solid curves) and $B(E2;7/2^-\rightarrow3/2^-)$
  (dashed curves) within the ground state rotational band.  (d)~The transition strength ratio
  $B(E2;5/2^-\rightarrow3/2^-)/B(E2;7/2^-\rightarrow3/2^-)$.  The Alaga ratio
  $12/5$ is shown for comparison (horizontal bar).  Calculated values are shown
  as functions of the basis parameter $\hw$, for $\Nmax=4$ to $10$ (as labeled).
\label{fig:convergence-9be}
}
\end{figure*}

The accuracy which can be obtained in solving the many-body problem is
limited by the highest $\Nmax$-truncated spaces which are computationally
accessible.  In $\isotope[9]{Be}$, an $\Nmax=13$ space (dimension
$\sim1.1\times10^{10}$) pushes the limits of current computational capabilities.
The dependence of calculated energies on the truncation $\Nmax$, as well as on
the basis parameter $\hw$, may be seen for the ground state $3/2^-_1$ (solid
curves) and ground state rotational band member $5/2^-_1$ (dashed curves) of
$\isotope[9]{Be}$ in Fig.~\ref{fig:convergence-9be}(a).
While the curves corresponding to succesive steps in $\Nmax$ are coming closer
together, and flattening in their dependence on $\hw$ as well, the calculated
energy eigenvalues are still changing by amounts on the order of $1\,\MeV$ with
each step in $\Nmax$, even near the variational minimum ($\hw\approx15\,\MeV$).
The use of softened interactions, such as the Daejeon16 interaction considered
here, ameliorates but clearly does not eliminate the challenge of convergence.

Given that the rotational spacings are also on the order of $\MeV$, it is not
$\textit{a priori}$ clear that detailed features of the rotational spectrum
should be well-resolved in the calculations.  Yet, despite limitations in
convergence, the rotational pattern of energy spacings is readily apparent.
While the energy eigenvalues of the individual states with a band may be
decreasing with $\Nmax$, the energies of states belonging to the same band
decrease together.

Returning to the lowest $3/2^-$ and $5/2^-$ states in $\isotope[9]{Be}$,
consider their energy difference, giving the excitation energy of the $5/2^-$
state within the band, shown in Fig.~\ref{fig:convergence-9be}(b).  This
difference flattens in $\hw$ in the vicinity of $\hw=15\,\MeV$, where it is changing
by $\lesssim0.05\,\MeV$ for each step in $\Nmax$.  We will note the degree of
convergence of relative energies within specific bands, as well as of the
excitation energies of bands relative to each other, as we explore the band
structure in more detail in the following sections.

For calculated $E2$ strengths, the convergence challenge is even more dramatic.
The dependences of the calculated $E2$ strengths, on $\Nmax$ and $\hw$, are
shown for the $5/2^-\rightarrow3/2^-$ (solid curves) and $7/2^-\rightarrow3/2^-$
transitions within the ground state band in Fig.~\ref{fig:convergence-9be}(c).
Neither set of curves, for either transition (solid and dashed curves,
respectively), is obviously approaching any particular stable, converged value.
Yet, the relative transition strengths among members of the same band are
already well-established at low $\Nmax$, with the ratios of transition strengths
approximately following Alaga rotational relations (see Figs.~6--8 and 17 of
Ref.~\cite{maris2015:berotor2} for quantitative analyses).  Returning to the
$5/2^-\rightarrow3/2^-$ and $7/2^-\rightarrow3/2^-$ transitions of
Fig.~\ref{fig:convergence-9be}(c), both families of curves have the same general
shape, differing rather in scale.  Taking the ratio of calculated values,
$B(E2;5/2\rightarrow3/2)/B(E2;7/2\rightarrow3/2)$, gives the values shown in
Fig.~\ref{fig:convergence-9be}(d).  While these are clearly not strictly
converged, there is a clear ``shoulder'' (inflection) in the $\hw$ dependence in
the vicinity of $\hw=15\,\MeV$, where the ratio is changing by $\lesssim5\%$ for
each step in $\Nmax$. The values may be compared to the Alaga ratio
$12/5\,(=2.4)$, from~(\ref{eqn:BE2-rotational}), for an ideal $K=3/2$ band.

Converged values for observables reflect an accurate solution of the many-body
problem as it has been mathematically formulated.  Whether or not the ensuing
values are in agreement with experiment is an entirely separate question.
Success depends not only on solving the many-body problem as it has been stated,
but on the fidelity of this many-body problem to the physical system in the
first place and, in particular, on the accuracy of the internucleon interaction
taken as input to the calculation.  That is, success also depends on the
structural integrity of the underlying layers in the tower of effective
theories.

While we attempt to provide some basic contact with experimental excitation
spectra for comparison in Figs.~\ref{fig:levels-9be}--\ref{fig:levels-7be}
(horizontal lines), the spin-parity assignments of many experimentally observed
levels are unknown or uncertain, and conflicting spin-parity assignments are
found in the literature.  For simplicity, the experimental levels shown in
Figs.~\ref{fig:levels-9be}--\ref{fig:levels-7be} are those assigned a unique
angular momentum and parity in the current experimental
evaluations~\cite{npa2002:005-007,npa2004:008-010,npa2012:011}, regardless of
whether or not this assignment is designated as tentative, with one
exception in $\isotope[11]{Be}$.\footnote{\fnexptassignment}

\section{$\isotope[9]{Be}$: $LS$ coupling scheme and Elliott rotation in the valence shell}
\label{sec:9be}

\subsection{Rotational spectrum and convergence}
\label{sec:9be:spectrum}
\begin{figure*}
\begin{center}
\includegraphics[width=\ifproofpre{0.90}{1}\hsize]{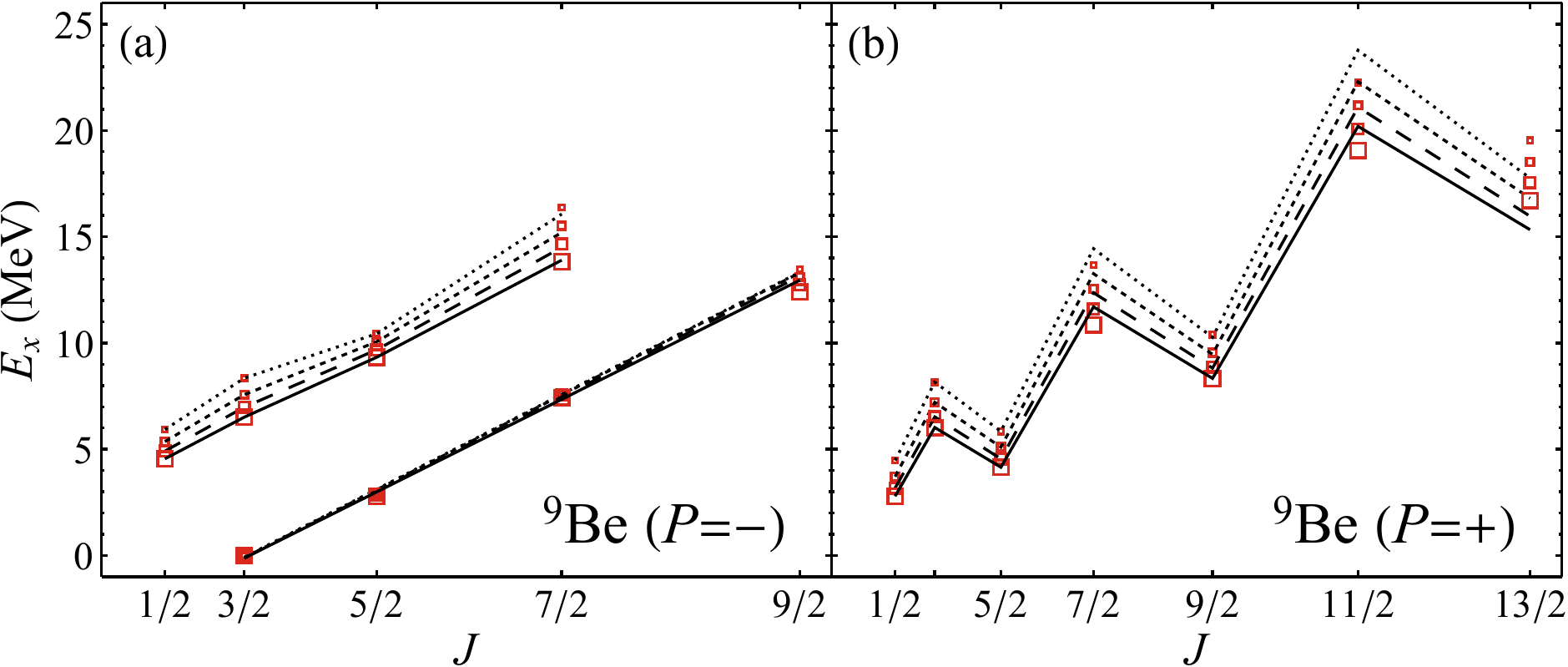}
\end{center}
\caption{Calculated energies for rotational band members in $\isotope[9]{Be}$,
  for (a)~negative and (b)~positive parity, shown as excitation energies
  relative to the negative parity ground state.  The calculated bands are for
  $\Nmax=4$ to $10$ or $\Nmax=5$ to $11$, respectively (dotted through solid
  curves).
\label{fig:band-energies-9be}
}
\end{figure*}

The rotational spectrum of $\isotope[9]{Be}$ provides a starting point for
illustrating many of the essential features of rotation, building on discussions
of the rotational structure of $\isotope[9]{Be}$ from prior NCCI
calculations~\cite{maris2015:berotor2,johnson2015:spin-orbit,caprio2015:berotor-ijmpe}.
The near-yrast states in the low-lying calculated spectrum of $\isotope[9]{Be}$
form three rotational bands: in the negative parity spectrum
[Fig.~\ref{fig:levels-9be}(a)], a $K^P=3/2^-$ ground state (yrast) band and
$K^P=1/2^-$ excited (yrare) band, and then, in the positive parity spectrum
[Fig.~\ref{fig:levels-9be}(b)], a $K^P=1/2^+$ yrast band.  These rotational
bands may be recognized from the energies, which approximately follow the
expected rotational energy spacings (with staggering for the $K=1/2$ bands), and
from the enhanced $E2$ connections between band members.

The main qualitative features of the rotational bands are robustly calculated,
that is, with only small residual dependences on the truncation of the
calculation.  The excitation energies of band members are shown for successive
values of $\Nmax$ in Fig.~\ref{fig:band-energies-9be}.  We may summarize the
general features, which are consistent with the earlier calculations with other
interactions (see Fig.~16 of Ref.~\cite{maris2015:berotor2}).  The excitation
energies of the two $K=1/2$ bands relative to the ground state are not as well
converged as the spacings within bands, but much better than the energy
eigenvalues themselves.  The $K^P=1/2^-$ excited band would appear to be rapidly
converging in excitation energy, towards an energy lower than that found at
$\Nmax=10$ but within $\sim1\,\MeV$~\cite{caprio2019:bebands-ntse18}.  The
staggering of energies within the calculated $K^P=1/2^-$ band decreases with
increasing $\Nmax$ and is consistent with zero ($a\lesssim0.1$)
[Fig.~\ref{fig:band-energies-9be}(a)].  The staggering within the $K^P=1/2^+$
band, in contrast, is pronounced ($a\approx 2$)
[Fig.~\ref{fig:band-energies-9be}(b)].

All three bands in the low-lying spectrum of $\isotope[9]{Be}$ terminate, and
they do so at angular momenta consistent with a simple $0\hw$ or $1\hw$ shell
model picture.  For negative parity, the maximal angular momentum which can be
constructed by coupling of the valence nucleons in the $p$ shell, in a $0\hw$
description of $\isotope[9]{Be}$, is $9/2$ [indicated by the vertical dashed
  line in Fig.~\ref{fig:levels-9be}(a)].  The $K^P=3/2^-$ band terminates at
$J=9/2$, while the $K^P=1/2^-$ band terminates at the lower angular momentum
$J=7/2$.  For positive parity, the maximal angular momentum which can be
generated in a $1\hw$ description of $\isotope[9]{Be}$, in particular, by
exciting one valence nucleon from the $p$ shell to the $sd$ shell, is $J=13/2$
[indicated by the vertical dashed line in Fig.~\ref{fig:levels-9be}(b)].  The
$K^P=1/2^+$ band likewise terminates at this angular
momentum.\footnote{\fnninebetermination}

The patterns of $E2$ transitions within the bands are generally consistent with
those expected from the Alaga rotational relations (Fig.~\ref{fig:rotor-e2}).
For instance, for the $K^P=1/2^+$ band, for which the transitions are most
clearly visible in the figure, due to the separation between lines afforded by
the energy staggering [Fig.~\ref{fig:levels-9be}(b)], we may compare to the
similar pattern for an ideal $K=1/2$ band (Fig.~\ref{fig:rotor-e2-network-k12}).

There is also a somewhat enhanced \textit{interband} transition between the
negative parity bands [Fig.~\ref{fig:levels-9be}(a)], from the $J=9/2$
terminating member of the ground state band to the $J=5/2$ member of the excited
band (noted also in Fig.~14 of Ref.~\cite{caprio2015:berotor-ijmpe}).  While
interband transitions are certainly possible in a rotational picture, these are
expected to follow Alaga rotational relations, in which all transition matrix
elements between the same two bands are proportional to a common intrinsic interband
transition matrix element~\cite{rowe2010:collective-motion}.  This single
enhanced $9/2^-_1\rightarrow5/2^-_2$ transition, without, say, a comparably
enhanced $7/2^-_1\rightarrow3/2^-_2$ transtion, is not expected in a simple
axially symmetric rotational picture.  Rather, it appears to represent a band
termination effect reflecting the limited dimension of the $0\hw$ shell model
space, which admits only one $J=9/2$ state, to be ``shared'' between the bands
(see also Sec.~\ref{sec:9be:am} below).

\subsection{Structure in oscillator space}
\label{sec:9be:oscillator}
\begin{figure}
\begin{center}
  \includegraphics[width=\ifproofpre{0.90}{0.50}\hsize]{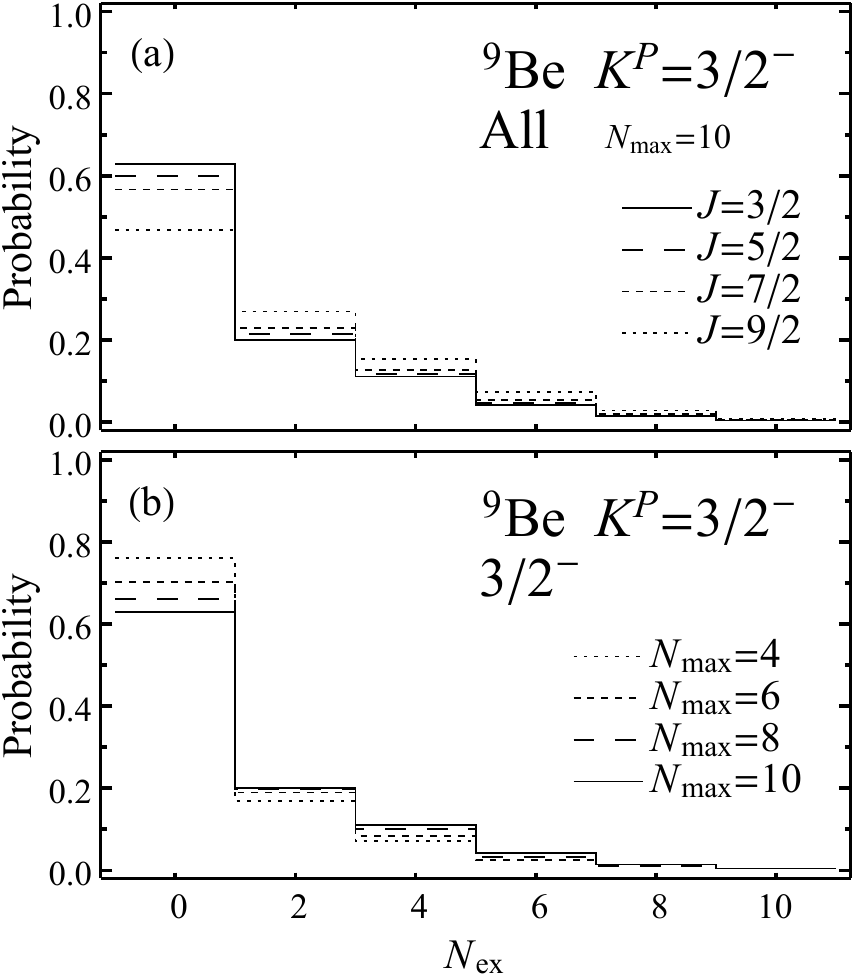}
\end{center}
\caption{Decompositions of $\isotope[9]{Be}$ $K^P=3/2^-$ ground state band
  members, with respect to number of excitation quanta $\Nex$ in the
  contributing oscillator configurations: (a)~Decompositions of all band members
  $3/2\leq J \leq 9/2$ (solid through dotted curves), as calculated for
  $\Nmax=10$.  (b)~Decomposition of the $3/2^-$ band head (\textit{i.e.}, the
  ground state), as calculated for $\Nmax=4$ to $10$ (dotted through solid
  curves).
\label{fig:decompositions-Nex-9be}
}
\end{figure}

A natural first question is then whether or not the rotational structure
presented by these states might have an effective description within a simple
valence shell model space ($0\hw$ or $1\hw$, for negative and positive parity,
respectively).  The calculated states in Fig.~\ref{fig:levels-9be}(a)
can be loosely identified as ``$0\hw$'' or ``$2\hw$'' (or higher $\Nex\hw$) states, in traditional shell model
terminology, based on their decompositions in the harmonic oscillator
basis.  Similarly the calculated states in Fig.~\ref{fig:levels-9be}(b)
can be loosely identified as ``$1\hw$'' or ``$3\hw$'' (or higher $\Nex\hw$).

The contributions from configurations of different $\Nex$ to the wave function
norm (or probability) are shown in Fig.~\ref{fig:decompositions-Nex-9be}, for
the calculated wave functions of the ground state band members.  The
contribution from $\Nex=0$ dominates in each band member
[Fig.~\ref{fig:decompositions-Nex-9be}(a)], although the details are dependent
upon the truncation of the calculation
[Fig.~\ref{fig:decompositions-Nex-9be}(b)].  In general, some of this
probability ``bleeds off'' to higher $\Nex$ as the wave functions are calculated
in higher $\Nmax$ spaces.  The rough classification of states as $0\hw$ or
$2\hw$ (shaded and open symbols, respectively) in
Figs.~\ref{fig:levels-9be}--\ref{fig:levels-7be} is determined simply by
considering whether the largest contribution to the wave function comes from
$\Nex=0$ oscillator many-body basis states or $\Nex=2$ oscillator many-body
basis states~\cite{mccoy2018:spncci-busteni17-URL,chen20xx:11be-xfer} (the
classification into $1\hw$ and $3\hw$ states for unnatural parity is
accomplished similarly).

Examination of the different oscillator-basis $\Nex$ contributions is
suggestive, and provides a potentially valuable diagnostic tool to recognize
qualitative patterns in the calculated spectra (as in the following
Secs.~\ref{sec:11be} and~\ref{sec:7be}).  However, it is important to keep in
mind that the $\Nex$ decomposition in an oscillator basis is at best an
approximate indicator of structure, even aside from the dependence on the
$\Nmax$ truncation of the calculation.  We may at most loosely identify
$\Nex\hw$ excitations in an oscillator basis with $\Nex \hw$ excitations in the
traditional shell model, which are taken to be particle-hole excitations above a
physically-meaningful mean-field (\textit{e.g.}, Hartree-Fock) vacuum.
Furthermore, even for fully converged calculations of the exact same wave
function, different results are obtained for the decomposition into oscillator
basis functions, depending upon the choice of length scale (or $\hw$ parameter)
for the oscillator basis into which the decomposition is being carried out
(here, recall, we are working with a basis parameter of $\hw=15\,\MeV$, chosen
near the variational energy minimum for calculations with the Daejeon16
interaction).

Nonetheless, for the calculated $\isotope[9]{Be}$ rotational band members, the
dominance of $\Nex=0$ contributions suggests that an effective description in
the valence shell may not be unreasonable.  Such an effective description could be
approached through, \textit{e.g.}, reformulation of the \textit{ab initio}
problem in a valence space obtained through application of the in-medium
similarity renormalization group (IM-SRG)~\cite{hergert2017:im-srg}.

\subsection{Angular momentum structure}
\label{sec:9be:am}
\begin{figure*}
\begin{center}
  \includegraphics[width=\ifproofpre{0.90}{1}\hsize]{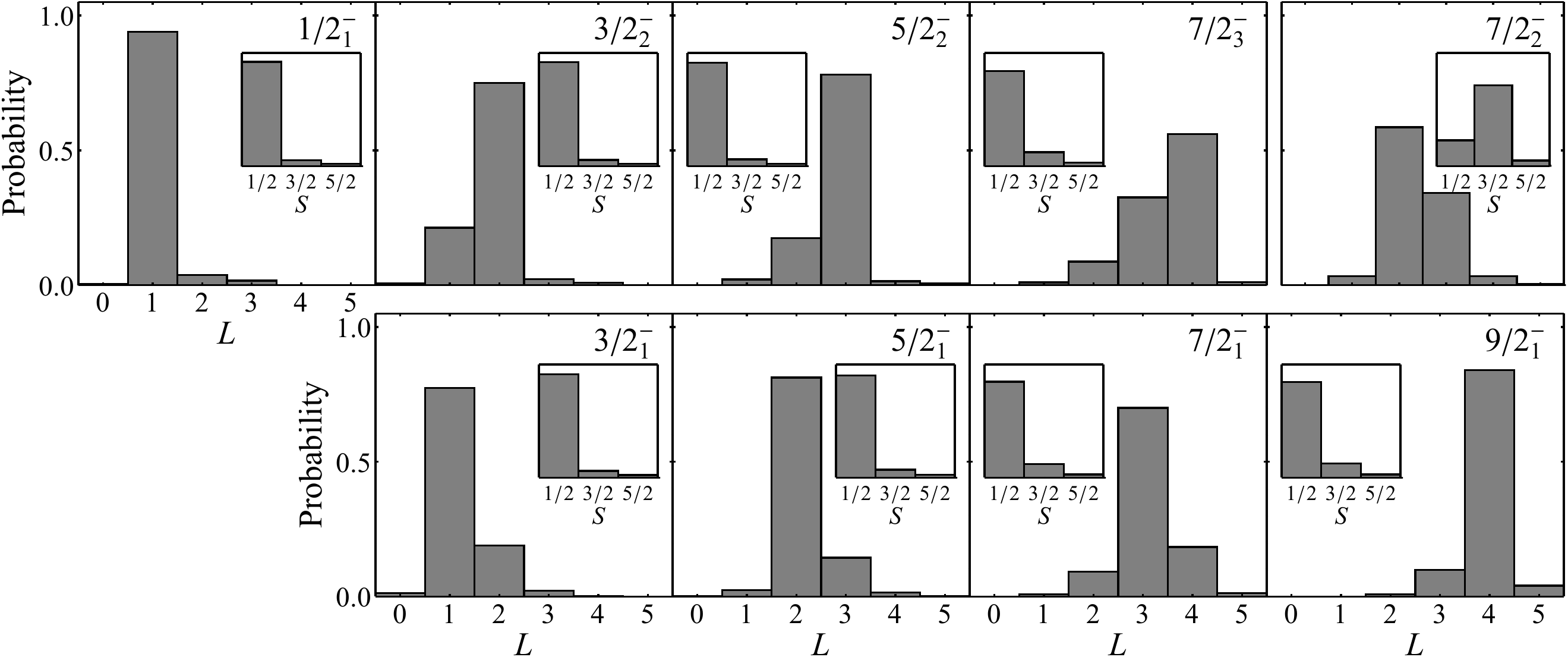}
\end{center}
\caption{Orbital angular momentum decompositions of $\isotope[9]{Be}$ negative
  parity states: $K^P=3/2^-$ ground state band members ($3/2^-_1$, $5/2^-_1$,
  $7/2^-_1$, $9/2^-_1$) (bottom), $K^P=1/2^-$ excited band members ($1/2^-_1$,
  $3/2^-_2$, $5/2^-_2$, and $7/2^-_3$) (top left), and an off-yrast state lying
  between the bands ($7/2^-_2$) (top right).  Spin angular momentum
  decompositions are shown as insets.  Based on wave functions calculated for
  $\Nmax=10$.
\label{fig:decompositions-am-9be}
}
\end{figure*}
\begin{figure}
\begin{center}
\includegraphics[width=\ifproofpre{0.90}{0.5}\hsize]{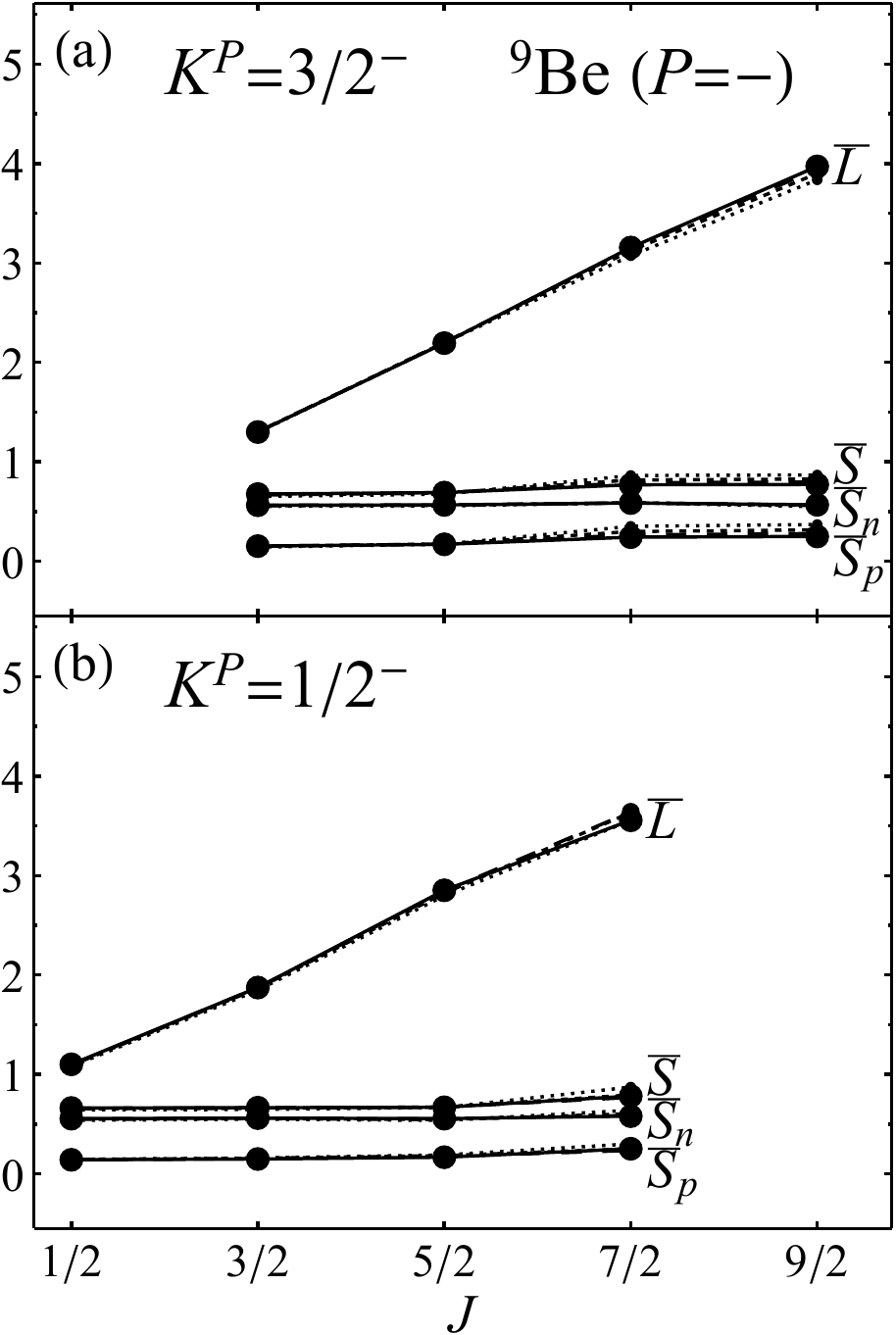}
\end{center}
\caption{Effective values of orbital and spin angular momenta for
  $\isotope[9]{Be}$ negative parity rotational band members: (a)~$K^P=3/2^-$
  ground state band and (b)~$K^P=1/2^-$ excited band.  Based on wave functions
  calculated for $\Nmax=4$ to $10$ (dotted through solid curves).
\label{fig:pmam-9be1}
}
\end{figure}
\begin{figure}
\begin{center}
\includegraphics[width=\ifproofpre{0.90}{0.5}\hsize]{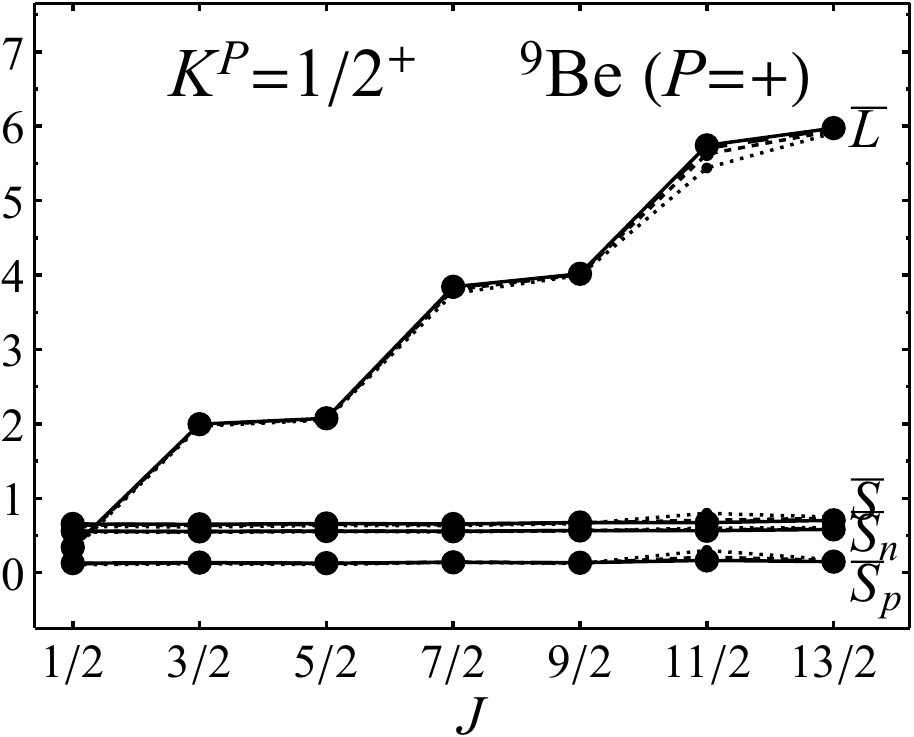}
\end{center}
\caption{Effective values of orbital and spin angular momenta for
  $\isotope[9]{Be}$ positive parity rotational band members ($K^P=1/2^+$ band).
  Based on wave functions calculated for $\Nmax=5$ to $11$ (dotted through solid
  curves).
\label{fig:pmam-9be0}
}
\end{figure}

The $LS$ angular momentum coupling scheme is understood to play a significant
role in the structure of $p$-shell nuclei, in possible competition with the $jj$
coupling scheme which becomes predominant in heavier
nuclei~\cite{inglis1953:p-shell,kurath1956:p-shell-intermediate}.  While $jj$ coupling refers to the role of
single-particle $j$ orbitals, the concept of $LS$ coupling is independent of
choice of single-particle basis, or even the concept of a 
single-particle basis.  It is defined rather in term of the total orbital (spatial)
angular momentum $\Lvec$ of the nucleons, and their total spin angular momentum $\Svec$, by the condition that the
many-body state have sharp angular momenta $L$ and $S$, combining to give the
total $J$.\footnote{\fnshellcoupling}

That the \textit{ab initio} calculated negative parity rotational band members
of $\isotope[9]{Be}$ largely obey $LS$ coupling was demonstrated in
Ref.~\cite{johnson2015:spin-orbit}, in the context of calculations with the
Entem-Machleidt \nthreelo{} interaction.  The portion of any NCCI calculated
wave function coming from contributions with a given $L$ or $S$ is not manifest
from its expansion in a tradititional $M$-scheme basis, but the so-called
``Lanczos trick''\footnote{\fnlanczostrick} may be used to decompose the
original calculated wave function into contributions from the different
eigenspaces of $\Lvec^2$ and $\Svec^2$, and thus according to $L$ and $S$.

For the $\isotope[9]{Be}$ negative parity band members, the $L$ and $S$ decompositions from the present
calcuations with the Daejeon16 interaction are
shown in Fig.~\ref{fig:decompositions-am-9be}.
The salient feature of the $LS$ structure,
discussed for the earlier calculations in
Refs.~\cite{johnson2015:spin-orbit,caprio2015:berotor-ijmpe}, is that the dominant $L$
contributions for successive band members are $L=1$, $2$, $3$, and $4$, while
the spin is predominantly $S=1/2$.  In the ground state $K^P=3/2^-$ band, the
angular momenta are coupled in the ``aligned'' sense, giving $J=L+1/2$, and thus
$J=3/2$, $5/2$, $7/2$, $9/2$.  In the excited $K^P=1/2^-$ band, these same angular momenta are coupled in the
``antialigned'' sense, giving $J=L-1/2$, and thus $J=1/2$, $3/2$, $5/2$, $7/2$.

A more concise overview of the angular momentum structure of these bands is
obtained by considering a single ``effective'' (or mean) orbital angular momentum
$\bar{L}$ for each state, defined in terms of the expectation value of the $\Lvec^2$
operator as\footnote{\fnpmam}
\begin{equation}
  \bar{L}(\bar{L}+1)\equiv\tbracket{\Lvec^2}.
\end{equation}
Effective proton spin, neutron spin, and spin angular
momenta ($\bar{S}_p$, $\bar{S}_n$, and $\bar{S}$, respectively) may be obtained
similarly, \textit{e.g.}, $\bar{S}(\bar{S}+1)\equiv\tbracket{\Svec^2}$.
Plotting these
quantities against $J$, as in Fig.~\ref{fig:pmam-9be1}, provides an
illuminating illustration of the linear growth in $L$ within both bands, along
with the near constant $S_p$, $S_n$, and $S$, and the shift between aligned and
antialigned coupling for the two bands.  Observe that the total spin
($S\approx1/2$) comes primarily from the neutron spin ($S_n\approx1/2$), with
the proton spins coupling to give a total near zero
($S_p\approx0$).\footnote{\fnamadmixture}

A natural simple interpretation, based on this $LS$ structure, identifies the
two negative parity bands in $\isotope[9]{Be}$ as $LS$ spin-flip partners,
involving the same orbital (that is, spatial) structure but opposite couplings
to the spin~\cite{caprio2015:berotor-ijmpe}.  The orbital motion is then
consistent with being rotational in nature, described by a $K_L=1$ band with
orbital angular momenta $L=1,\ldots,4$.  That is, the orbital motion is based on
an intrinsic state with projection $K_L=1$ of the orbital angular momentum along
the symmetry axis.  In this limit of the rotational interpretation, based on
weak coupling of spin to spatial rotation,\footnote{\fncollectivecoupling} the
total angular momenta $J=1/2,3/2^2,5/2^2,7/2^2,9/2$ then follow simply by
angular momentum coupling of $L$ and $S$.

For the positive parity band ($K^P=1/2^+$), a decidedly different $LS$ angular
momentum structure is obtained, as may be seen
from the effective angular momenta in Fig.~\ref{fig:pmam-9be0}.  Here,
the spin is again predominantly $S\approx 1/2$ and again arises from the neutrons, but the
orbital angular momenta now form a stair-step pattern: $L\approx 0,2,2,4,4,6,6$,
for the $J=1/2,\ldots,13/2$ states, respectively.  A natural simple description
for this band is thus based on orbital motion consisting of a $K_L=0$ rotational
band, containing even values of angular momentum ($L=0,\ldots,6$).  Then,
successive band members alternate between anti-aligned and aligned couplings of
$L$ and $S$.

We have already noted that the negative parity band members in the \textit{ab
  initio} calculations appear to be primarily $0\hw$ states
(Sec.~\ref{sec:9be:oscillator}) and could thus potentially be described within a
shell-model effective theory.  It is thus informative to compare their $LS$
momentum structure with that expected in an Elliott $\grpsu{3}$ shell model
description.  The states which are brought lowest in energy by an $\grpsu{3}$
quadrupole-quadrupole Hamiltonian are those forming the leading irrep of
$\grpsu{3}$ arising in the valence shell model space of the nucleus,
\textit{i.e.}, the irrep having the largest eigenvalue for the $\grpsu{3}$
Casimir operator.  For $\isotope[9]{Be}$, the leading irrep has quantum numbers
$(\lambda,\mu)=(3,1)$ and, from fermionic antisymmetry constraints, occurs in
association with a total spin $S=1/2$.  According to the
$\grpsu{3}\rightarrow\grpso{3}$ angular momentum branching rule, the $(3,1)$
irrep indeed contains a single $K_L=1$ band, with $L=1,2,3,4$, exactly as found
here in the \textit{ab initio} calculations.

While we have focused thus far on the strict $LS$ weak coupling limit (in the
both the shell model and collective model senses of weak coupling), such does
not provide a satisfactory description beyond reproducing the overall angular
momentum $J$ content of the bands.  Indeed, the simple picture of a $K_L$ band,
with rotation only
in the orbital degrees of freedom, would give an energy spacing proportional to
$L(L+1)$ rather than $J(J+1)$, as seen in Fig.~\ref{fig:levels-9be}, and sets
aside the question of the Coriolis staggering.

However, as demonstrated by Elliott and Wilsdon~\cite{elliott1968:su3-part4} in
the $\grpsu{3}$ picture, a modest spin-orbit interaction (intermediate coupling,
in the shell model sense) serves to mix the states of different $L$ but coupled
with spin to give the same final $J$ (\textit{i.e.}, $L=J\pm1/2$) arising from
the two spin-flip partner bands or, more generally, within the same $\grpsu{3}$
irrep.  The resulting spectrum consists of states which approximate a
conventional rotational spectrum, with states of definite $K=K_L+K_S$ (strong
coupling, in the collective rotational sense) as in~(\ref{eqn:psi}) and
following the rotational energy formula~(\ref{eqn:E-rotational-no-coriolis})
or~(\ref{eqn:E-rotational-coriolis}).  The resulting $\grpsu{3}$ interpretation
of the $\isotope[9]{Be}$ ground state band, including mixing, is discussed by
Millener~\cite{millener2001:light-nuclei}.  Certainly, some such $L$ mixing
($L=J\pm1/2$) is apparent for the calculated states in
Fig.~\ref{fig:decompositions-am-9be}.

While weak coupling (good $L$) states form an orthogonal basis, the strong
rotational coupling (good $K$) states obtained within a shell model space, in
the Elliott-Wilsdon picture, lead to a breakdown of the band structure at high
$J$, \textit{i.e.}, near band termination. The resulting state in the
$\grpsu{3}$ shell model picture cannot be uniquely identified with a specific
$K$ band (recall the ambiguous transition pattern from the terminating $9/2^-$
state in Sec.~\ref{sec:9be:spectrum}).

The orbital angular momentum projections $K_L$ for the rotational intrinsic
state suggested by these observations on the $LS$ structure of $\isotope[9]{Be}$
are also consistent with the appealing intuitive description of these bands as
arising from nuclear molecular rotation.  In a molecular description,
$\isotope[9]{Be}$ is composed of two $\alpha$ clusters plus a single valence
neutron, which occupies a molecular orbital in the potential generated by the
clusters~\cite{blair1958:alpha-model-9be-inelastic,kunz1960:alpha-model-light,hiura1963:alpha-model-9be,dellarocca2018:cluster-shell-model-part1-9be-9b}.
Within each $\alpha$ particle, the spins of both protons and of both neutrons
must couple to give zero resultant spin, leaving only the spin contribution of
the last neutron ($S=1/2$).  Thus, while the rotational band members in the
negative parity space could be consistent with such a picture, other low-lying
states, \textit{e.g.}, the calculated $7/2^-_2$ state, with dominant spin
contribution $S=3/2$ [Fig.~\ref{fig:decompositions-am-9be} (top right)], must
involve breaking of an $\alpha$ particle.

In the phenomenological cluster molecular orbital description, as discussed in
Ref.~\cite{dellarocca2018:cluster-shell-model-part1-9be-9b}, both the
$K^P=3/2^-$ and $K^P=1/2^-$ negative parity bands are obtained from an intrinsic
state in which the neutron is in a $\pi$ orbital, \textit{i.e.}, giving angular
momentum projection $K_L=1$ along the symmetry axis defined by the clusters.
Rotational strong-coupling intrinsic states with definite $K=3/2$ and $K=1/2$
then arise from the aligned and anti-aligned combinations $K=K_L\pm K_S$,
respectively, with the projection $K_S=1/2$ of the neutron spin along the
symmetry axis.  The positive parity $K^P=1/2^+$ band instead arises from an
intrinsic state for which the neutron is in a $\sigma$ orbital, \textit{i.e.},
giving $K_L=0$, and thus $K=K_S=1/2$.

\section{$\isotope[11]{Be}$: Rotation outside the valence space}
\label{sec:11be}

\subsection{Rotational spectrum and convergence}
\label{sec:11be:spectrum}
\begin{figure*}
\begin{center}
\includegraphics[width=\ifproofpre{0.90}{1}\hsize]{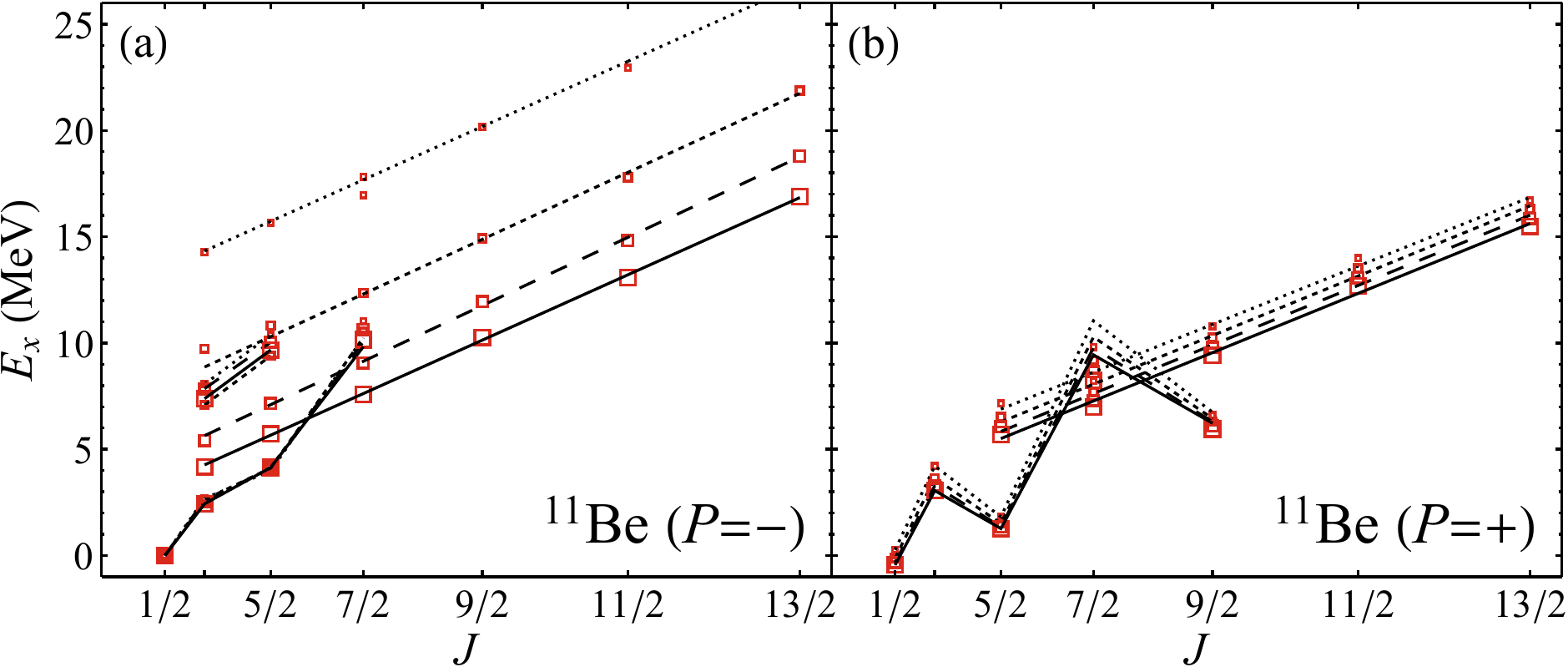}
\end{center}
\caption{Calculated energies for rotational band members in $\isotope[11]{Be}$,
  for (a)~negative and (b)~positive parity, shown as excitation energies
  relative to the negative parity ground state.  The calculated bands are for
  $\Nmax=4$ to $10$ or $\Nmax=5$ to $11$, respectively (dotted through solid curves).
\label{fig:band-energies-11be}
}
\end{figure*}

The low-lying calculated spectrum of $\isotope[11]{Be}$, shown in
Fig.~\ref{fig:levels-11be}, brings in new rotational characteristics, most
notably, a natural parity rotational band which lies outside the realm of a
$0\hw$ effective theory.  These rotational bands again have well-defined angular
momentum structures in an $LS$ coupling picture.

The nucleus $\isotope[11]{Be}$ is well-known for the so-called parity inversion
which arises in the spectrum.  That is, the experimental $1/2^+$ ground state is
of positive (and thus unnatural) parity, contrary to what might naively be
expected from shell-model considerations~\cite{talmi1960:11be-shell-inversion}.
It lies $0.320\,\MeV$ below the $1/2^-$ lowest negative (natural) parity
level~\cite{npa2012:011}.  This parity inversion, which is generally considered
challenging to reproduce in an \textit{ab initio} NCCI framework and known to be
sensitive to the details of the
interaction~\cite{forssen2005:ncsm-9be-11be,calci2016:11be-inversion-ncsmc}, is
obtained in calculations with the Daejeon16
interaction~\cite{kim:daejeon16-ntse18-INPRESS}.  Considering the near
degeneracy of the lowest $0\hw$ and $1\hw$ states, within a shell-model
interpretation, it is perhaps not surprising that $2\hw$ states should also be
found at low excitation energy.
\begin{figure*}[t]  
\begin{center}
  \includegraphics[width=\ifproofpre{0.90}{1}\hsize]{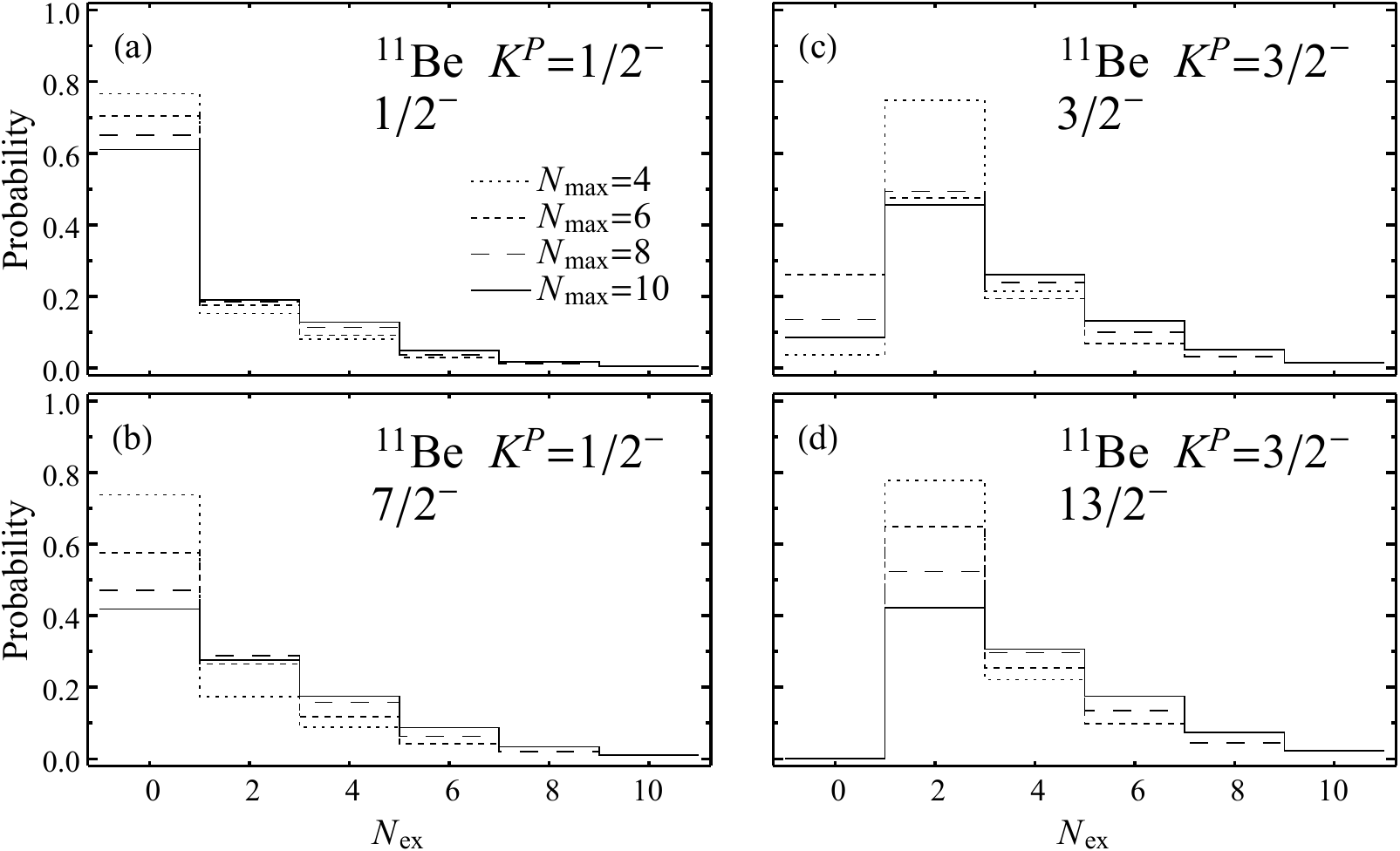}
\end{center}
\caption{Decompositions of representative $\isotope[11]{Be}$ negative parity
  rotational band members: (Left)~The $K^P=1/2^-$ negative parity ground state
  band's (a)~$1/2^-$ band head and (b)~$7/2^-$ terminating state.  (Right)~The
  $K^P=3/2^-$ negative parity long band's (c)~$3/2^-$ band head and (d)~$13/2^-$
  terminating state.  Decompositions are calculated from wave functions
  obtained for $\Nmax=4$ to $10$ (dotted through solid curves).
\label{fig:decompositions-Nex-11be}
}
\end{figure*}

Starting with the negative parity spectrum, the calculated $1/2^-_1$, $3/2^-_1$,
$5/2^-_1$, and $7/2^-_2$ states [Fig.~\ref{fig:levels-11be}(a)] may be
identified as forming a $K^P=1/2^-$ band.  The angular momenta in the negative
parity ground state band extend to the maximal angular momentum $J=7/2$ which
can be constructed in the valence ($0\hw$) space.  The $E2$ transitions follow
the characteristic transition pattern for a $K=1/2$ band
(Fig.~\ref{fig:rotor-e2-network-k12}).  There is modest positive Coriolis staggering
($a\approx0.4$).  We shall refer to this band, in the following discussion, as
the ``negative parity ground state band'', in that it is built on the ground
state of the negative parity space, in distinction to the ``positive parity
ground state band'', built on the ground state of the positive parity space
(which becomes the overall ground state both in the high $\Nmax$ calculations
here and in experiment).

However, there are also enhanced $E2$ connections from the negative parity
ground state band to the $3/2^-_3$ and $5/2^-_3$ states. These states are
themselves connected by a strong $E2$ transition, comparable to the in-band
transitions, suggesting that these states could be described as constituting a
$K^P=3/2^-$ band, albeit a very short one, which would spectroscopically be described
as a side band to the negative parity ground state band.

Then, threading between these bands in energy [Fig.~\ref{fig:levels-11be}(a)],
is another $K^P=3/2^-$ band.  This band becomes yrast at $J=7/2$, and then
extends through this maximal valence angular momentum with no noticeable
disruption to the rotational energies, finally terminating at $J=13/2$.  In
comparison with the ``short'' negative parity ground state band ($A\approx 0.6
\,\MeV$), this ``long'' band has energies which follow a line with a
significantly shallower slope ($A\approx 0.3\,\MeV$).  Between the short band
and the long band, there is thus an approximate doubling of the moment of
inertia.  The long band does not have significant $E2$ connections to the short
bands.

Considerations of convergence are especially important for the $K^P=3/2^-$ long
band.  The excitation energies of the negative parity band members at various
$\Nmax$ are traced out in Fig.~\ref{fig:band-energies-11be}.  The long band
starts at high excitation energy in the spectrum, at low $\Nmax$, but rapidly
descends with increasing $\Nmax$.  For instance, the $7/2^-$ band member only
becomes yrast at $\Nmax=8$.  The low final energy for the long band in the
present calculations with the Daejeon16 interaction, at the highest $\Nmax$
considered here ($\Nmax=10$), reflects the comparatively rapid convergence
obtained with the Daejeon16 interaction.  In comparison, in calculations with
the JISP16 or \nnloopt{} interactions, at this same $\Nmax$, the band head energy
still lies well above
$10\,\MeV$~\cite{maris2015:berotor2,caprio2015:berotor-ijmpe,caprio2019:bebands-ntse18}.
Without attempting any detailed extrapolation here, we can observe that the
calculated energies of the long band members appear to be converging towards
those of a corresponding experimentally identified excited rotational
band~\cite{vonoertzen1997:be-alpha-rotational,bohlen2008:be-band} (see
Refs.~\cite{caprio2019:bebands-ntse18,chen20xx:11be-xfer} for further discussion
of and comparison with the experimental levels and rotational energy parameters).

Within the positive parity spectrum [Fig.~\ref{fig:levels-11be}(b)], the yrast
and near-yrast states can likewise be identified as forming rotational bands,
based on energies and enhanced $E2$ connections.  The $K^P=1/2^+$ positive
parity ground state band terminates at $J=9/2$, and exhibits large positive
Coriolis staggering ($a\approx1.9$).  (Note the excellent agreement of the
\textit{ab initio} predicted excitation energies within the band for the lowest
three band members with experimentally observed levels.)  Next lies an excited
$K^P=5/2^+$ band, which becomes yrast at $J=11/2$ and terminates at $J=13/2$,
the maximal angular momentum accessible in the $1\hw$ space.  The moments of
inertia of these bands differ (the slope parameter is $A\approx0.4\,\MeV$ for
the $K^P=1/2^+$ band and $A\approx0.25\,\MeV$ for the $K^P=5/2^+$ band).

There are some modestly enhanced $E2$ transistions between these bands and to
other low-lying states (namely, $3/2^+_2$ and $5/2^+_3$).  Most noticeable,
though, is the $E2$ transition pattern from the terminating $13/2^+$ state,
which is at the maximal angular momentum accessible in the $1\hw$ space.
Although the energy of this state is roughly consistent with membership in the
$K^P=5/2^+$ band, the strongest $E2$ transition is to the $K^P=1/2^+$ band,
suggesting a breakdown of the rotational strong coupling picture at high $J$
(recall the termination effects from Sec.~\ref{sec:9be:am}).

\begin{figure*}
\begin{center}
  \includegraphics[width=\ifproofpre{0.90}{1}\hsize]{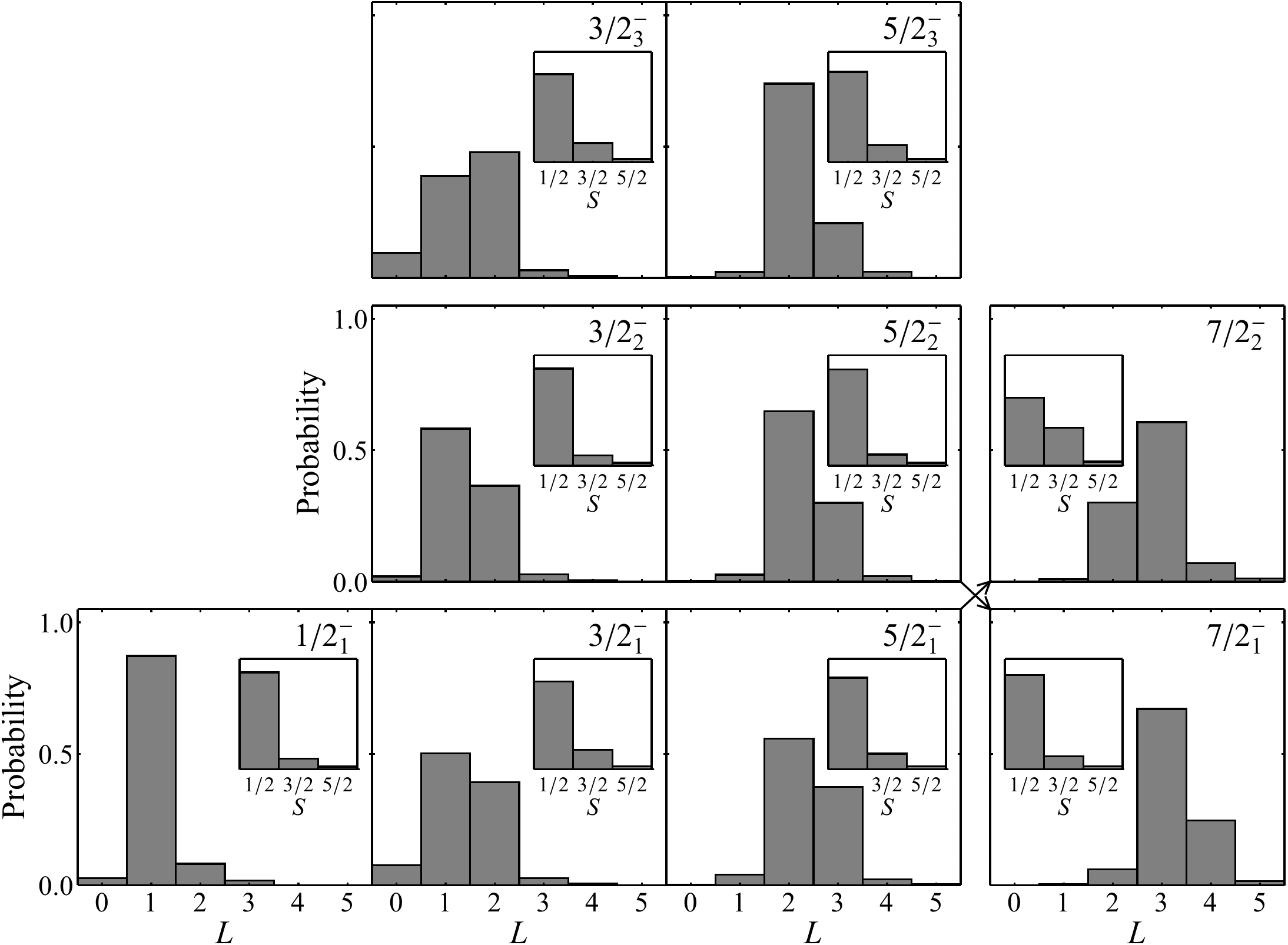}
\end{center}
\caption{Orbital angular momentum decompositions of $\isotope[11]{Be}$ negative
  parity states: $K^P=1/2^-$ negative parity ground state band members
  ($1/2^-_1$, $3/2^-_1$, $5/2^-_1$, and $7/2^-_2$) (bottom), $K^P=3/2^-$ long
  band members ($3/2^-_2$, $5/2^-_2$, and $7/2^-_1$) (middle), and $K^P=3/2^-$
  side band members ($3/2^-_3$ and $5/2^-_3$) (top).  The crossed arrows
  indicate the change in energy ordering of the first two bands between $J=5/2$
  and $7/2$.  Spin angular momentum decompositions are shown as insets.  Based
  on wave functions calculated for $\Nmax=10$.
\label{fig:decompositions-am-11be}
}
\end{figure*}
\begin{figure}
\begin{center}
\includegraphics[width=\ifproofpre{0.90}{0.5}\hsize]{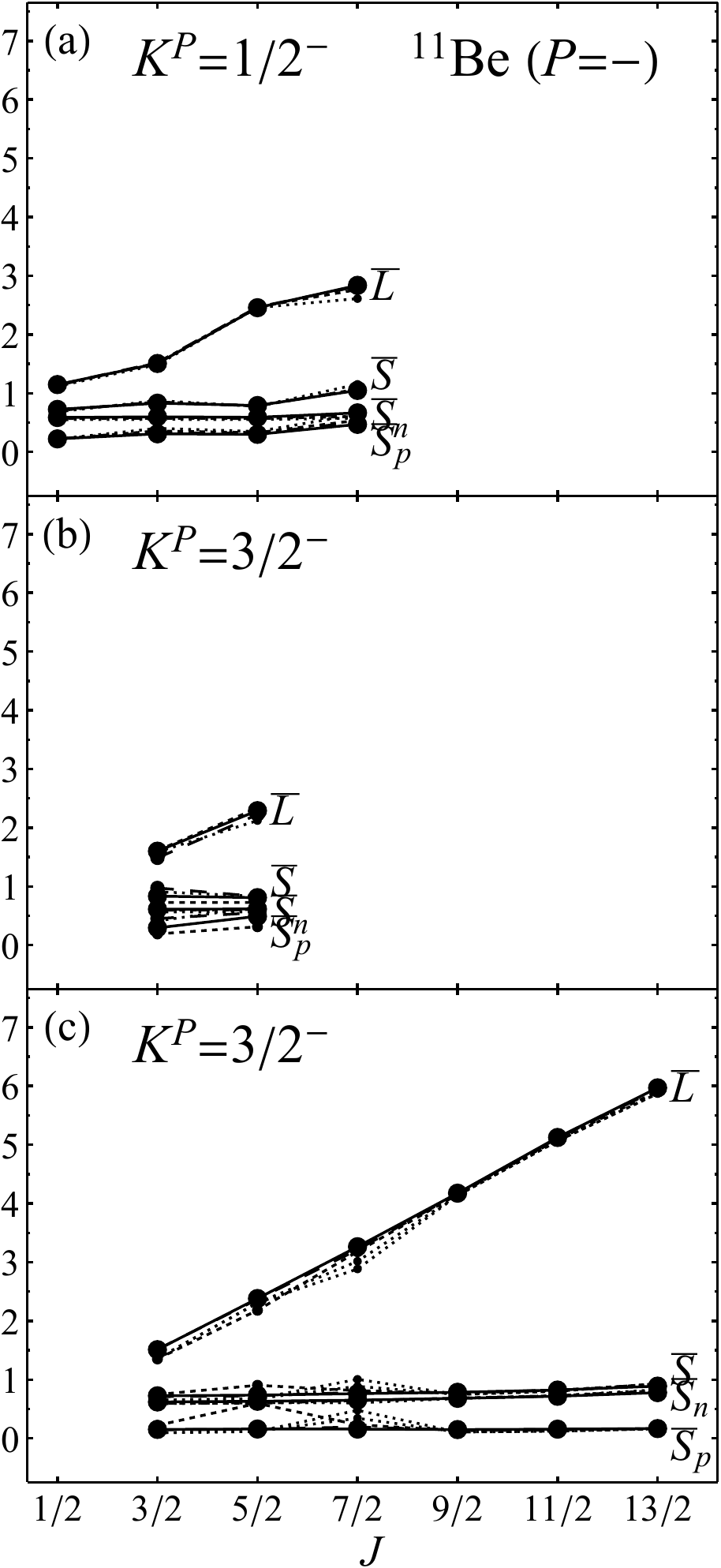}
\end{center}
\caption{Effective values of angular momenta for $\isotope[11]{Be}$ negative parity
  rotational band members: (a)~$K^P=1/2^-$ negative parity ground state band,
  (b)~$K^P=3/2^-$ negative parity side band, and (c)~$K^P=3/2^-$ negative parity long band.
  Based on wave functions calculated for $\Nmax=4$ to $10$
  (dotted through solid curves).
\label{fig:pmam-11be1}
}
\end{figure}
\begin{figure}
\begin{center}
\includegraphics[width=\ifproofpre{0.90}{0.5}\hsize]{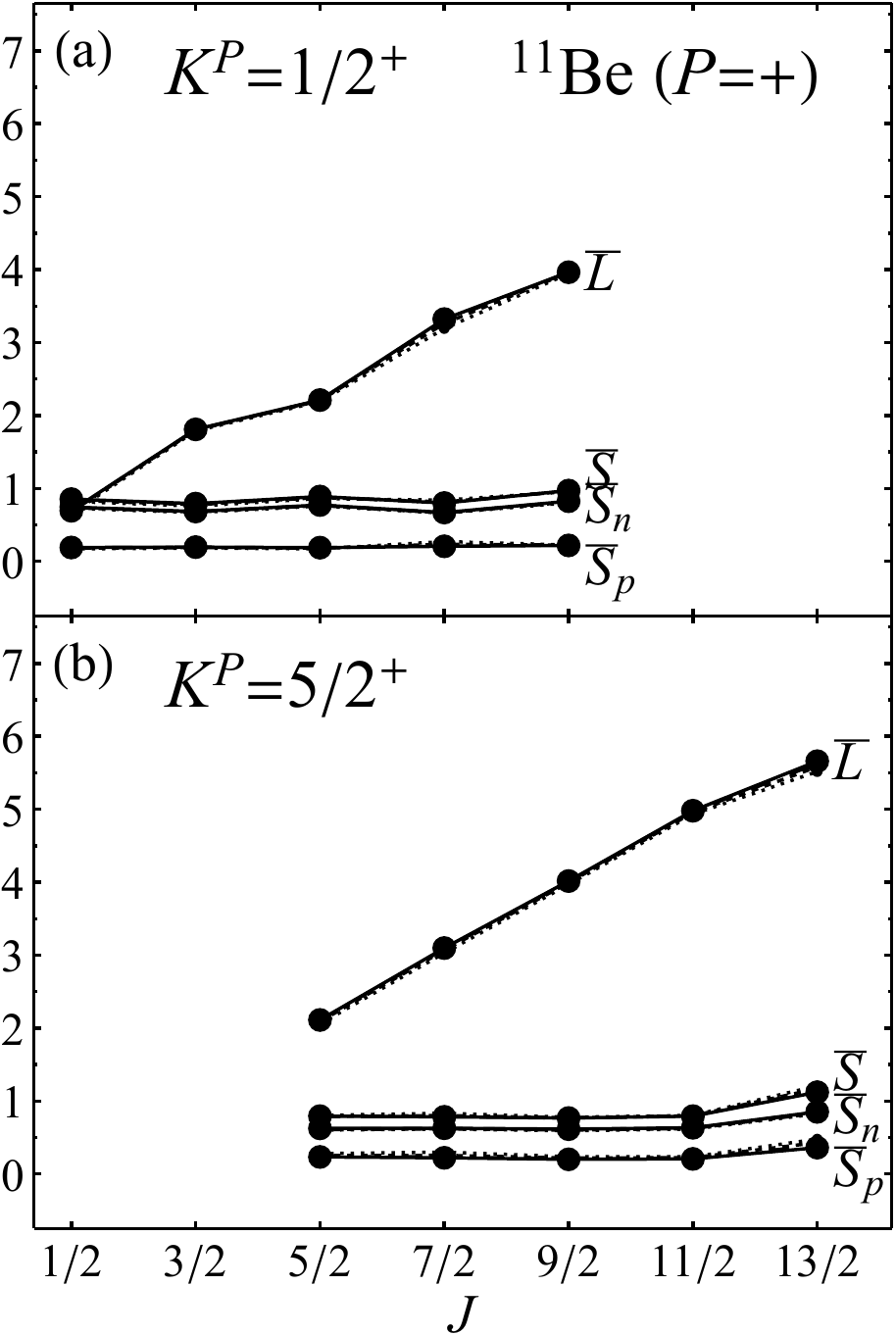}
\end{center}
\caption{Effective values of angular momenta for $\isotope[11]{Be}$ positive parity
  rotational band members: 
  (a)~$K^P=1/2^+$ positive parity ground state band and (b)~$K^P=5/2^+$
  positive parity excited band.  Based on wave functions calculated for $\Nmax=5$ to $11$
  (dotted through solid curves).
\label{fig:pmam-11be0}
}
\end{figure}

\subsection{Structure in oscillator space}
\label{sec:11be:oscillator}

The levels in the negative parity ground state band have largest contributions
coming from $0\hw$ oscillator configurations, as may be seen for representative
band members in Fig.~\ref{fig:decompositions-Nex-11be}(a,b).  The members of
the $K^P=3/2^-$ side band have similar decompositions.

However, for the states constituting the $K^P=3/2^-$ long band, a qualitatively
different oscillator decomposition is obtained
[Fig.~\ref{fig:decompositions-Nex-11be}(c,d)].  While band members with angular
momenta above the maximal valence angular momentum ($J=7/2$) cannot receive any
contribution from $0\hw$ oscillator basis configurations
[Fig.~\ref{fig:decompositions-Nex-11be}(d)], even for those band members
lying beneath the maximum valence angular momentum the $0\hw$
contribution is highly suppressed [Fig.~\ref{fig:decompositions-Nex-11be}(c)],
as initially noted in Ref.~\cite{chen20xx:11be-xfer}.  The largest contribution comes
from $2\hw$ basis states, after which the probability distribution falls off
gradually for higher $\Nex$.

\subsection{Angular momentum structure}
\label{sec:11be:am}

The states making up the rotational bands in $\isotope[11]{Be}$, like those in
$\isotope[9]{Be}$, again follow comparatively simple patterns when viewed in
terms of an $LS$ coupling scheme, as indicated by their $LS$ decompositions, as
shown in Fig.~\ref{fig:decompositions-am-11be}, or, again, more simply in terms
of the evolution of the effective $\bar{L}$ and $\bar{S}$ values as functions of
$J$ within the band, as shown in Figs.~\ref{fig:pmam-11be1}
and~\ref{fig:pmam-11be0}.
For all these bands in $\isotope[11]{Be}$, the 
predominant total spin $S=1/2$ arises from the neutrons ($S_n=1/2$), while the
proton spin vanishes ($S_p=0$).

For the $K^P=3/2^-$ long band [Fig.~\ref{fig:decompositions-am-11be} (middle)],
the angular momentum structure is comparatively straightforward.  The
$J=3/2,\ldots,13/2$ band members have orbital angular momenta of predominantly
$L=1,\ldots,6$, respectively.  The simple linear growth in $L$ with $J$ may be
seen most clearly in Fig.~\ref{fig:pmam-11be1}(c).  Thus, the orbital motion is
consistent with a band built on an $K_L=1$ state for intrinsic motion, which
then combines with the total spin $S=1/2$ in aligned coupling to give $J=L+1/2$.

For the $K^P=1/2^-$ negative parity ground state band and $K^P=3/2^-$ side band,
if we simply examine the effective $\bar{L}$ values
[Fig.~\ref{fig:pmam-11be1}(a,b)], the pattern is less obvious.  However, from
the detailed angular momentum decompositions
[Fig.~\ref{fig:decompositions-am-11be} (bottom,top)], it becomes clear that the
contributing orbital angular momenta are $L=1,2,3$.  The total set of angular
momenta ($J=1/2,3/2^2,5/2^2,7/2$) constituting the negative parity ground state
band and its side band are consistent with a single $K_L=1$
rotational band in the orbital motion, with $L=1,2,3$, combining with the spin
in antialigned ($J=1/2,3/2,5/2$) and aligned ($J=3/2,5/2,7/2$) couplings.
However, here there is clearly much stronger mixing between the states of same
$J$ but different $L$ than in the two negative parity bands in
$\isotope[9]{Be}$, for which the angular momentum decompositions more cleanly
indicate the aligned and antialigned couplings of a single $K_L=1$ band
with $S=1/2$.

Suggestively, the angular momentum structure for these negative parity ground
state and side bands is again exactly as expected from a simple Elliott
$\grpsu{3}$ picture, as described by Millener~\cite{millener2001:light-nuclei}.
For $\isotope[11]{Be}$, the leading irrep has $(\lambda,\mu)=(2,1)$ and arises with
$S=1/2$.  By the $\grpsu{3}\rightarrow\grpso{3}$ angular momentum branching
rule, this irrep indeed contains a single $K_L=1$ band with $L=1,2,3$, giving rise to
$K=1/2$ and $3/2$ bands.

For the positive parity ground state band ($K^P=1/2^+$)
[Fig.~\ref{fig:pmam-11be0}(a)], the orbital motion is described by a $K_L=0$ band
comprised of even angular momenta, much as for the postive parity band of
$\isotope[9]{Be}$, but here terminating at $L=4$ ($L=0,2,4$).  These orbital
angular momenta again couple alternately in antialigned and aligned couplings
with the spin $S=1/2$ to give $J=1/2,\ldots,9/2$.

Then, for the positive parity excited band ($K^P=5/2^+$)
[Fig.~\ref{fig:pmam-11be0}(b)], we now seem to find a $K_L=2$ orbital motion,
with $L=2,3,4,5,6$.  These orbital angular momenta combine in aligned coupling
with the spin ($S=1/2$) to give $J=5/2,\ldots,13/2$.  (Ostensibly a band arising
from the antialigned coupling might be found at higher excitation energy.)

Note that the spin structure ($S=1/2$ from neutrons) found in the calculated
rotational bands is again consistent with alpha cluster molecular structure.  In
the molecular description of $\isotope[11]{Be}$, the proton spins are coupled
pairwise to zero, within alpha particles, as are the spins of the neutrons
within the alpha particles.  The total spin thus arises from the three valence
neutrons, and ostensibly just the last unpaired valence neutron.

It is then perhaps reassuring that organization of the structure into rotational
bands for the orbital motion, with the $K_L$ values suggested by the present
interpretation of the \textit{ab initio} calculations, is generally consistent
with the description obtained in antisymmetrized molecular dynamics (AMD)
calculations for $\isotope[11]{Be}$, in terms of cluster molecular orbitals (see
Sec.~3.1.2 of Ref.~\cite{kanadaenyo2012:amd-cluster}).  That is, the negative
parity ground state band ($K^P=1/2^-$) is based on a $\pi^3$ configuration, and
the long band ($K^P=3/2^-$) is based on a $\pi\sigma^2$ configuration,
consistent with $K_L=1$ from an unpaired neutron in a $\pi$ orbital.  Then, the
lowest positive parity band ($K=1/2$) in the AMD description is based on a
$\pi^2\sigma$ configuration, consistent with $K_L=0$ from destructive addition
of $K_L=\pm1$ contributions from the two paired $\pi$ orbitals (it is also
natural to obtain $K_L=2$, from constructive addition, as found here for the
$K^P=5/2^+$ excited band).

\section{$\isotope[7]{Be}$: Quadrupole excitation}
\label{sec:7be}

\subsection{Rotational spectrum and discussion}
\label{sec:7be:spectrum}
\begin{figure}
\begin{center}
\includegraphics[width=\ifproofpre{0.90}{0.5}\hsize]{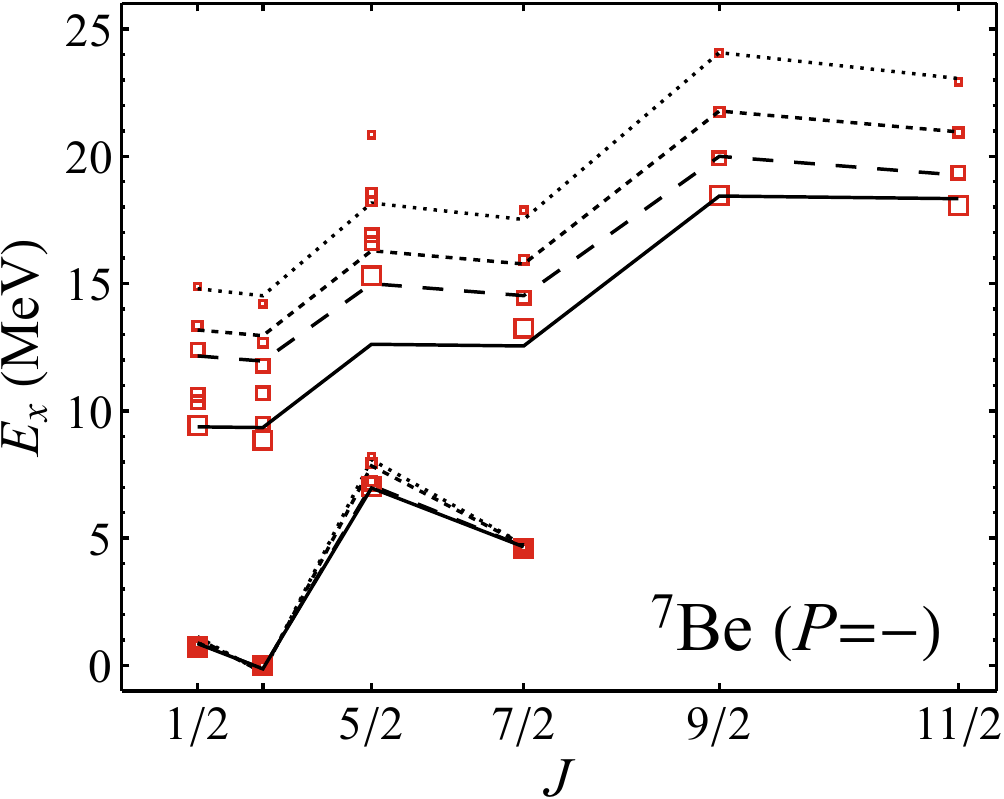}
\end{center}
\caption{Calculated energies for $\isotope[7]{Be}$ rotational band members,
  shown as excitation energies relative to the negative parity ground state.
  The calculated bands are for $\Nmax=8$ to $14$ (dotted through solid curves).
\label{fig:band-energies-7be}
}
\end{figure}
\begin{figure}
\begin{center}
  \includegraphics[width=\ifproofpre{0.90}{0.50}\hsize]{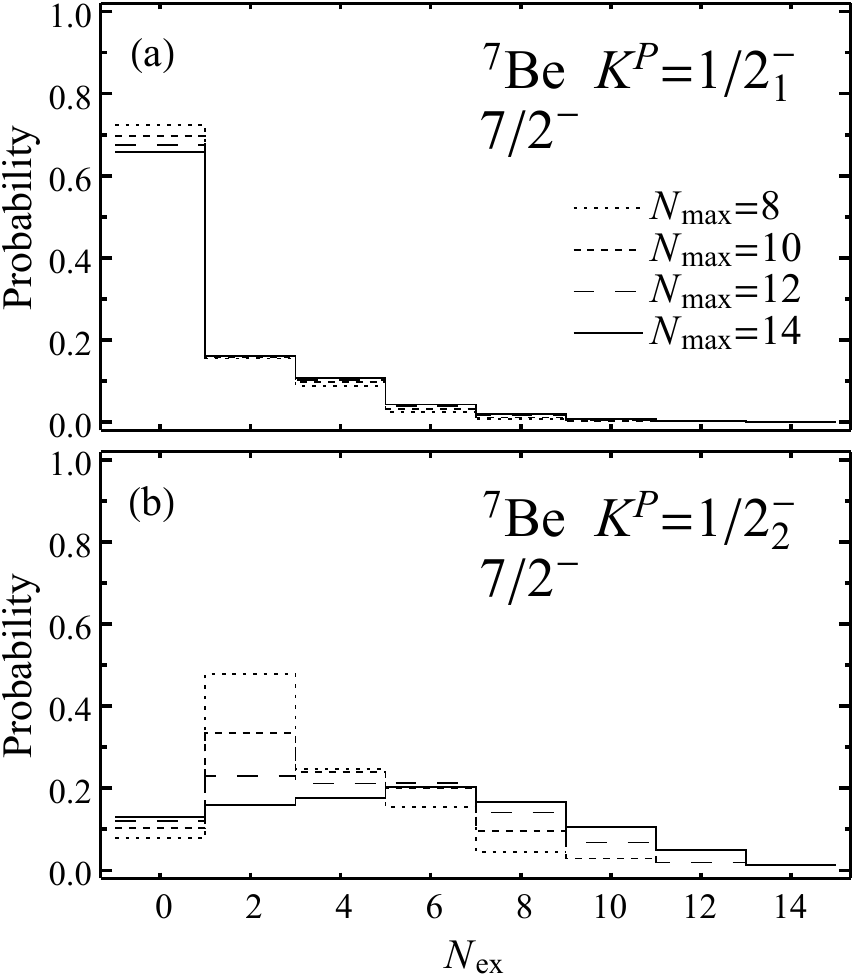}
\end{center}
\caption{Decompositions of representative $\isotope[7]{Be}$ negative parity
  rotational band members: (a)~the $K^P=1/2^-$ ground state band's $7/2^-$
  terminating state and (b)~the $K^P=1/2^-$ excited band's $7/2^-$ member.
  Decompositions are calculated from wave functions obtained for $\Nmax=8$ to
  $14$ (dotted through solid curves).
\label{fig:decompositions-Nex-7be}
}
\end{figure}
\begin{figure}
\begin{center}
\includegraphics[width=\ifproofpre{0.90}{0.5}\hsize]{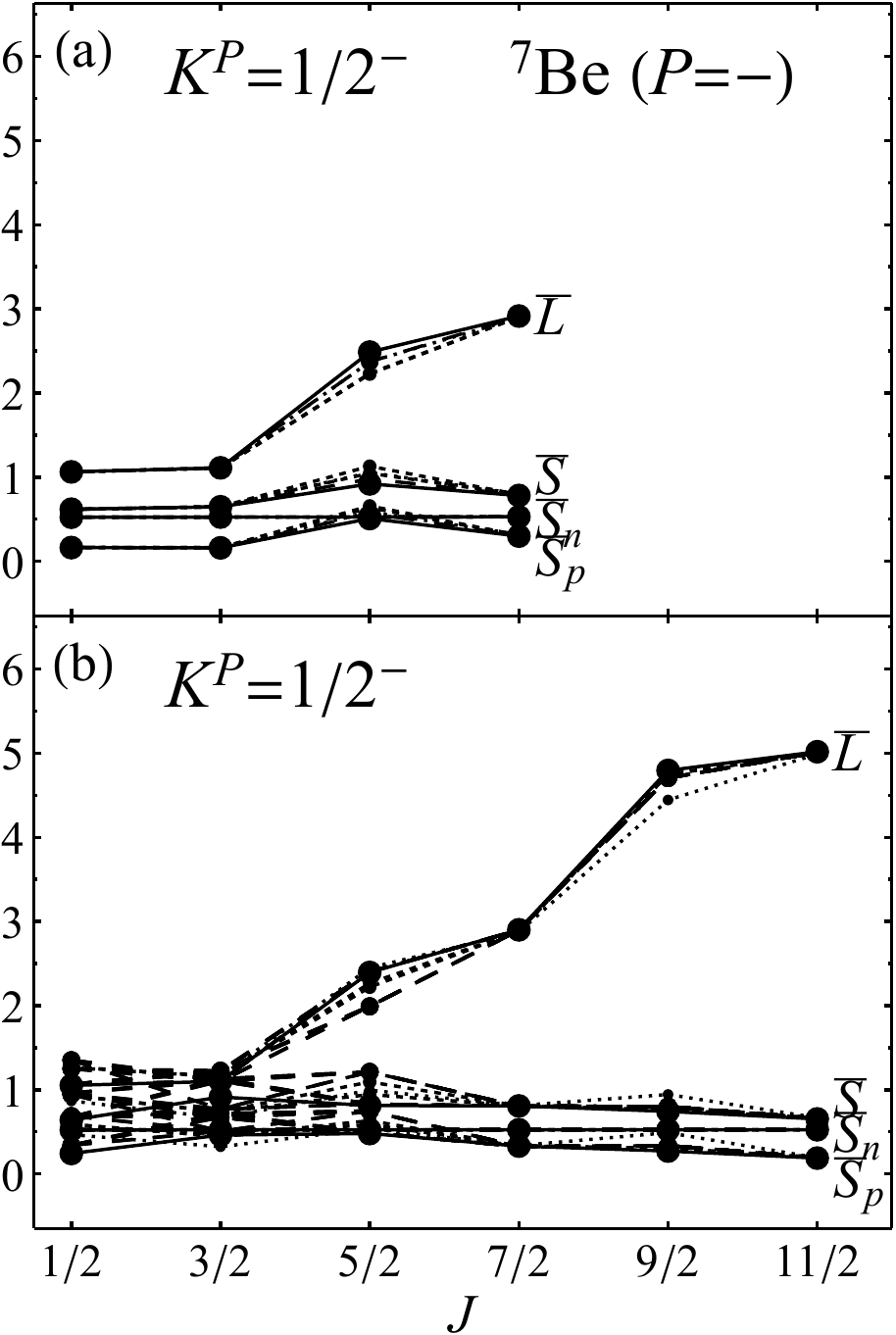}
\end{center}
\caption{Effective values of orbital and spin angular momenta ($\bar{L}$,
  $\bar{S}_p$, $\bar{S}_n$, and $\bar{S}$, as labeled) for $\isotope[7]{Be}$
  rotational band members: (a)~$K^P=1/2^-$ ground state band and (b)~$K^P=1/2^-$
  excited band.  Based on wave functions calculated for $\Nmax=8$ to $14$
  (dotted through solid curves).
\label{fig:pmam-7be1}
}
\end{figure}

The rotational structure emerging in the \textit{ab initio} calculations for
$\isotope[7]{Be}$, as shown in Fig.~\ref{fig:levels-7be}, shares some of the
characteristics we have just explored for $\isotope[9]{Be}$ and
$\isotope[11]{Be}$.  However, it also hints at the emergence of further
effective degrees of freedom.  In particular, it demonstrates the relevance of
quadrupole degrees of freedom and indicates a richer role for dynamical symmetry
as an organizing scheme for collective excitations.

The calculated negative parity spectrum (Fig.~\ref{fig:levels-7be}) contains a
$K^P=1/2^-$ ground state band, which terminates the maximal valence angular
momentum ($J=7/2$) for $\isotope[7]{Be}$.  The staggering of energies in this
band is of the type obtained for a negative value of the Coriolis decoupling
parameter $a$ (that is, the $J=1/2,5/2,\ldots$ members are raised and
$J=3/2,7/2,\ldots$ members lowered, contrary to the other $K=1/2$ bands
discussed above).  Indeed, this staggering is sufficiently pronounced that the
$3/2^-$ band member becomes the ground state, both in the calculations and in
experiment.

The $5/2^-$ band member, which is staggered upwards in energy, forms a close
doublet with another, apparently non-rotational $5/2^-$ state.  As the
calculated energies evolve with the basis truncation $\Nmax$, the $5/2^-$ member
of the ground state band begins slightly above the non-rotational $5/2^-$ state,
in low-$\Nmax$ calculations, and then ends up at lower energy, in high-$\Nmax$
calculations.  The crossing occurs at $\Nmax=10$, at which point the $E2$
strengths indicate significant two-state mixing.  Note that a close $5/2^-$
doublet is observed experimentally as well (Fig.~\ref{fig:levels-7be}).  (With
the exception of this $5/2^-$ band member, the calculated excitation energies
within the ground state band are seen to be essentially independent of $\Nmax$
in Fig.~\ref{fig:band-energies-7be}.)

The calculated wave functions for the ground state band members are
predominantly $0\hw$ in character, as illustrated for the terminating $7/2^-$
band member in Fig.~\ref{fig:decompositions-Nex-7be}(a).  Moreover, inspection
of the angular momenta, in Fig.~\ref{fig:pmam-7be1}(a), indicates once again a
straightforward angular momentum structure in the $LS$ scheme.  The spin is
predominantly $S\approx 1/2$ (and arises from the neutrons).  The orbital
angular momenta again form a stair-step pattern, as for the positive parity
$K=1/2$ band in $\isotope[9]{Be}$ (recall Fig.~\ref{fig:pmam-9be0}), but now
with odd values for $L$, namely, $L= 1,1,3,3$, for the $J=1/2,\ldots,7/2$
states, respectively.  The angular momentum content of the $5/2^-$ band member
is perturbed by two-state mixing with the nearby non-rotational $5/2^-$ state,
which has major contributions with $S_p=1$ and $S=3/2$, especially at their
closest approach in energy ($\Nmax=10$).

Due to basic symmetry considerations, a $K=0$ rotational band may contain either
only even angular momenta or only odd angular momenta, depending whether the
intrinsic wave function takes on a positive or negative sign ($r=\pm1$) under a
rotation of $\pi$ around an axis perpendicular to the symmetry axis,
respectively~\cite{rogers1965:nonspherical-nuclei,rowe2010:collective-motion}.
Thus, the angular momentum structure of the ground state band in
$\isotope[7]{Be}$ is consistent with an orbital motion described by a $K_L=0$,
$r=-1$ rotational band, containing odd values of angular momentum ($L=1,3$),
which then combine alternately in antialigned and aligned coupling with the spin
($S=1/2$) to give successive band members.

Such a picture has, in fact, long been speculated from a cluster molecular
orbital description, in which $\isotope[7]{Be}$ may be viewed as two $\alpha$
particles plus a neutron hole occupying a $\sigma$ molecular orbital.  See the
discussion (of the mirror nuclide $\isotope[7]{Li}$) in Sec.~10 of
Ref.~\cite{inglis1953:p-shell}, where it is, in particular, suggested that the
ground state $3/2^-$ and first excited $1/2^-$ states are obtained from the
aligned and antialigned couplings, respectively, of the same $K_L=0$, $L=1$
motion to spin ($S=1/2$).  In the limit of large cluster separations, this
molecular orbital description reduces to removal of a neutron from one $\alpha$
to form an $\alpha$-$\isotope[3]{He}$ molecule (here it may be helpful to refer
to a molecular orbital diagram, as in Fig.~5 of
Ref.~\cite{dellarocca2018:cluster-shell-model-part1-9be-9b}).

If we now look beyond the maximal valence angular momentum in
Fig.~\ref{fig:levels-7be}, there is a puzzling feature to the spectrum.  One
particular $9/2^-$ state, slightly above the yrast line, and the yrast $11/2^-$
state have strong $\Delta J=2$ $E2$ transitions to the $5/2^-$ and $7/2^-$
ground state band members, respectively.  Indeed, these transitions are
comparable in strength to in-band transitions.

On the one hand, such enhanced transitions could suggest that these $9/2^-$ and
$11/2^-$ states might be taken as possible ground state band members, as
discussed in
Refs.~\cite{caprio2013:berotor,maris2015:berotor2,caprio2015:berotor-ijmpe}. The
calculated energies lie high relative to what we would expect for $J=9/2$ and
$11/2$ band members, based on the rotional energy
formula~(\ref{eqn:E-rotational-coriolis}), but the excitation energies are not
well converged and are still decreasing with $\Nmax$, so such a comparison is
ambiguous.

On the other hand, these $9/2^-$ and $11/2^-$ states also have $\Delta J=2$
transitions, of comparable strength, to specific higher-lying $5/2^-$ and
$7/2^-$ states, respectively.  Indeed, in the present calculations, we can trace
out a complete excited $K^P=1/2^-$ band from the $E2$ strengths, terminating
with these yrast $9/2^-$ and $11/2^-$ states
(Fig.~\ref{fig:levels-7be}).\footnote{\fnsevenbehint}  Thus, the transitions from
these $9/2^-$ and $11/2^-$ states to the ground state band members are not
in-band transitions, but rather highly enhanced interband $E2$ transitions.

In a calculation at any given $\Nmax$, such as the $\Nmax=14$ calculation in
Fig.~\ref{fig:levels-7be}, individual band members, especially the
upward-staggered states ($J=1/2,5/2,9/2$), are subject to transient two-state
mixing with other ``background'' states, due to accidental degeneracies in
energy (see Sec.~IV\,A of Ref.~\cite{maris2015:berotor2}).  This complication
will tend to somewhat obscure any investigation of the properties of the band
members, but the basic features are apparent.

In particular, the excited band members, whether above the maximal valence
angular momentum or below, have their largest contributions coming from $2\hw$
or higher oscillator contributions, as shown in
Fig.~\ref{fig:decompositions-Nex-7be}(b).  At modest $\Nmax$ values, the
decomposition is sharply peaked at $2\hw$, but the distribution becomes much
broader for the highest $\Nmax$ calculations, peaking at $4\hw$ or even $6\hw$
contributions.

Then, the orbital angular momenta, in Fig.~\ref{fig:pmam-7be1}(b), follow the
same stair-step pattern as for the ground state band, consistent with a $K_L=0$
rotational motion restricted to odd $L$ ($r=-1$), but now extending to $L=5$.
(The upward-staggered $J=1/2,5/2,9/2$ band members, and the lower $J$ band
members in general, are more subject to transient contamination of their angular
momentum content from mixing at specific $\Nmax$ values.)

Although transient mixing or fragmentation makes it difficult to accurately
track the convergence of the excitation energies of the lower-$J$ band members
with increasing $\Nmax$, this convergence is indicated to the extent possible in
Fig.~\ref{fig:band-energies-7be}.  At least superficially, the excited
$K^P=1/2^-$ band in $\isotope[7]{Be}$ would seem to resemble the long
$K^P=3/2^-$ band in $\isotope[11]{Be}$, which similarly receives predominant
contributions from $2\hw$ or higher configurations and extends beyond the
maximal valence angular momentum (Sec.~\ref{sec:11be:spectrum}).  The calculated
excitation energies for both bands move rapidly downward with increasing $\Nmax$
[compare Fig.~\ref{fig:band-energies-11be}(a) for $\isotope[11]{Be}$ and
  Fig.~\ref{fig:band-energies-7be} for $\isotope[7]{Be}$].

However, there are also notable differences in the convergence properties.  In
particular, the long band in $\isotope[11]{Be}$ seems to be rapidly approaching
a final, converged excitation energy (with each successive step in $\Nmax$, the
change in excitation energy is smaller by about half).  In contrast, the
convergence pattern of the energies of the excited band in $\isotope[7]{Be}$ is
still not clearly defined.

\subsection{Dynamical symmetry structure}
\label{sec:7be:dynamical}

To understand the relationship between the ground state band and the excited
band in the \textit{ab initio} calculations for $\isotope[7]{Be}$, and how this
might hint at the emergence of new effective theories, we turn to the sympletic
group $\grpsptr$ in three
dimensions~\cite{rosensteel1977:sp6r-shell,rosensteel1980:sp6r-shell,rowe1985:micro-collective-sp6r}.
This group augments the generators of Elliott's $\grpu{3}$, which all conserve
the total number of oscillator quanta, with further generators which physically
represent the creation and annihilation operators for giant monopole and
quadrupole resonances.  These latter generators either create or destroy two
oscillator quanta and can therefore connect $0\hw$ and $2\hw$ states.  The giant
quadrupole resonance operators in particular are naturally implicated if $0\hw$
and $2\hw$ states are connected by strong $E2$ transitions.

We are aided by calculations carried out in a symplectic no-core configuration
interaction (SpNCCI)
framework~\cite{mccoy2018:spncci-busteni17-URL,mccoy2018:diss,mccoy2019:spncci-ntse18},
in which the nuclear many-body basis is organized into irreps of the group chain
\begin{multline}
\label{eqn:sp-chain}
\bigl[
  \underset{\sigma}{\grpsptr}
  \supset
  \underset{\omega}{\grpu{3}}
  \supset
  \underset{L}{\grpso{3}}
\bigr]
\times
\underset{S}{\grpsu[S]{2}}
\supset\underset{J}{\grpsu[J]{2}},
\end{multline}
with quantum numbers as shown, where
$\sigma\equiv\Nsex(\lambda_\sigma,\mu_\sigma)$ and
$\omega\equiv\Nwex(\lambda_\omega,\mu_\omega)$.  It is then straightforward to
extract the decomposition of the calculated wave functions not only with respect
to the Elliott $\grpu{3}$ quantum numbers $\Nwex(\lambda_\omega,\mu_\omega)$, but the $\grpsptr$ quantum numbers
$\Nsex(\lambda_\sigma,\mu_\sigma)$ as well.  By way of explanation, we
simply note here that a single $\grpsptr$ irrep is obtained by starting from
some ``lowest'' $\grpu{3}$ irrep $\sigma$, \textit{i.e.}, having the lowest
number of oscillator excitation quanta $\Nsex$ within this particular $\grpsptr$
irrep.  Then the $\grpsptr$ irrep consists of an infinite tower of $\grpu{3}$ irreps
$\omega$, \textit{i.e.}, with $\Nwex=\Nsex,\Nsex+2,\ldots$, obtained by
laddering repeatedly with the giant resonance creation operators.

From calculations of $\isotope[7]{Be}$ in a more restricted space ($\Nmax=6$),
described in Refs.~\cite{mccoy2018:diss,mccoyXXXX:spfamilies}, the $\grpu{3}$
structure of the bands becomes clear.  The ground state band members are
identified with the $0\hw$ $\grpsu{3}$ irrep $(3,0)$, with $S=1/2$, while the
excited band members are identified with the $2\hw$ $\grpsu{3}$ irrep $(5,0)$,
again with $S=1/2$.  The $\grpsu{3}\rightarrow\grpso{3}$ angular momentum
branching rule indeed yields that a $(3,0)$ irrep is comprised of a single
$K_L=0$ band with $L=1,3$, while a $(5,0)$ irrep is comprised of a single
$K_L=0$ band with $L=1,3,5$.  Thus, the $\Nex$ and angular momentum structure
detailed above for the rotational bands follows simply from an Elliott effective
description.

Yet, it is also found that the $\grpu{3}$ irrep describing the excited band has
a particular symplectic structure.  The $\grpu{3}$ and spin quantum numbers
$\omega S\equiv \Nwex(\lambda_\omega,\mu_\omega) S=2(5,0) 1/2$ are far from
unique in the space for $\isotope[7]{Be}$, comprising a subspace of dimension
$12$.  One (and only one) particular linear combination within this
$12$-dimensional space yields the $\grpu{3}$ irrep which is a member of the
$\sigma S=\Nsex(\lambda_\sigma,\mu_\sigma) S=0(3,0)1/2$ symplectic irrep, that
is, the symplectic irrep built on the ground state band's $\grpu{3}$ irrep by
laddering with the giant resonance operators.  And it turns out that the
calculated excited band members lie within this symplectic irrep, at the level
of $\sim50\%$--$80\%$ of their norm.

Thus, the ground state and excited bands would seem to represent an example of
approximate $\grpsptr\supset\grpu{3}$ dynamical symmetry.  The strong interband
$E2$ transitions reflect their connection by the giant quadrupole operator,
which, as a generator of $\grpsptr$, acts entirely within an $\grpsptr$ irrep.

The dynamical symmetry relationship between the bands is suggestive of an
emerging physical structure.  In the contraction limit, obtained for large
values of the quantum numbers, the microscopic symplectic picture gives way to a
collective interpretation of the dynamics in terms of effective coupled
rotational and vibrational (giant resonance) degrees of
freedom~\cite{leblanc1984:symplectic-rotor-vibrator,rowe1985:micro-collective-sp6r,leblanc1986:symplectic-quadrupole,rowe2016:micsmacs}.

In such an extremely light and minimally bound nucleus as $\isotope[7]{Be}$, the
physical interpretation is less clear.  The contraction limit, essentially a
semiclassical interpretation, is likely not well realized, and, when working
within a bound state formalism as we are here, possible interactions with
the scattering continuum could alter the interpretation of computed states.
Nonetheless, at the very least, the emergence of rotational bands connected by
quadrupole excitations in a symplectic dynamical symmetry scheme in these
\textit{ab initio} calculations may be taken as a possible harbinger of emergent
rotational-vibrational structure in heavier and more strongly bound systems.

\section{Conclusion}
\label{sec:concl}

Microscopic \textit{ab initio} theory offers the potential of predictive power,
by addressing the nuclear many-body problem in its full glory, without
presupposing mean-field structure, collective degrees of freedom, or many-body
dynamical symmetries.  It may therefore not be immediately obvious how to
extract, from the results of such large-scale calculations, a physically
intuitive understanding of the nucleus, of the kind afforded by models defined
in terms of effective degrees of freedom.  As has been said in the (slightly
different) context of large-scale shell model calculations, ``even if such
calculations are possible using high-speed computers, the results are difficult
to interpret physically and the consquences of agreement or disagreement with
the data are much less intuitively informative\textellipsis nor could one even
begin to understand the resultant wave functions''~\cite{casten2000:ns}.

Yet, \textit{ab initio} theory has advanced to the point where the resulting
calculated spectra do exhibit signatures of such emergent phenomena, including
rotational features substantially resembling those observed in experiment.  We
see this in the present examples taken from the odd-mass $\isotope{Be}$ isotopes
(Figs.~\ref{fig:levels-9be}--\ref{fig:levels-7be}).

On one hand, most obviously, the calculated spectra may be taken in the spirit
of a numerical experiment, permitting access to a rich set of observables in
these nuclei, such as $E2$
strengths~\cite{mccutchan2007:10be-dsam-gfmc,mccutchan2012:10c-dsam,datar2013:8be-radiative,henderson2019:7be-coulex},
which would otherwise be largely inaccessible due to experimental limitations.
The traditional phenomenological analyis then takes over, starting from the
computed ``data'', permitting the identification of emergent structural
features.  This is how collective features in the spectrum, such as rotational
bands, are first identified.

On the other hand, the mere fact that the computed wave functions are ``large''
in an \textit{ab initio} approach (for NCCI calculations, say, comprised of
amplitudes for $\lesssim 10^{10}$ basis configurations) is in itself not an
insurmountable impediment to discerning simple structure within these wave
functions.  The same large-scale computational tools which are used to generate
the wave functions are also available to assist in their judicious analysis.

We have focused here on simple shell structure, which is hinted at by
decompositions in oscillator space and band termination phenomena, and on simple
angular momentum structure, which is apparent from $LS$
decompositions~\cite{johnson2015:spin-orbit}.  These basic observations are
consistent with and suggestive of a more complete understanding in terms of a
richer structure of Elliott $\grpsu{3}$ and symplectic dynamical symmetries, as
indicated both by the spectroscopy and by group theoretical decompositions of
the wave
functions~\cite{dytrych2007:sp-ncsm-evidence,dytrych2007:sp-ncsm-dominance,draayer2012:sa-ncsm-qghn11,dytrych2013:su3ncsm,launey2016:sa-nscm,mccoy2018:spncci-busteni17-URL,mccoy2018:diss,mccoyXXXX:spfamilies}.
A complementary understanding~\cite{suzuki1986:sp6r-alpha-cluster-me} likely
comes in terms of cluster molecular
structure~\cite{neff2004:cluster-fmd,kanadaenyo2012:amd-cluster}.  For these
light nuclei, where only a handful of lowest-energy states are truly bound, a
more complete understanding requires going beyond a bound-state formalism
towards approaches which can more directly identify clustering degrees of
freedom underlying
resonances~\cite{kravvaris2017:ab-initio-cluster-8be-10be-12c,vorabbi2019:7be-7li-ncsmc}.

From these observations, we have provided some insight into the links between
microscopic theory and emergent effective degrees of freedom, by recognizing
basic structures and patterns in the \textit{ab initio} results.  The analyses
presented here are essentially simple, focusing on spectroscopy and a few basic
decompositions of the wave functions, but they already indicate the emergence of
mean field structure, $LS$ or intermediate-coupling rotation, and a nascent
giant quadrupole degree of freedom.


\begin{acknowledgements}
We thank Calvin W.~Johnson and Tom\'{a}\v{s} Dytrych for valuable
discussions. This material is based upon work supported by the U.S.~Department
of Energy, Office of Science, under Award Numbers DE-FG02-95ER-40934,
DESC00018223 (SciDAC4/NUCLEI), and DE-FG02-87ER40371, and by the U.S.~National
Science Foundation under Award Number NSF-PHY05-52843.  TRIUMF receives federal
funding via a contribution agreement with the National Research Council of
Canada.  This research used computational resources of the University of Notre
Dame Center for Research Computing and of the National Energy Research
Scientific Computing Center (NERSC), a U.S.~Department of Energy, Office of
Science, user facility supported under Contract~DE-AC02-05CH11231.
\end{acknowledgements}





\end{document}